\documentclass[journal]{IEEEtran}
\usepackage{amsmath,amsfonts,amssymb}
\usepackage{algorithmic}
\usepackage{algorithm}
\usepackage{array}
\usepackage{booktabs}
\usepackage{multirow}
\usepackage[table,xcdraw]{xcolor}
\usepackage{makecell}
\usepackage{url}
\usepackage{verbatim}
\usepackage{graphicx}
\usepackage[caption=false,font=footnotesize]{subfig}
\usepackage{textcomp}
\usepackage{hhline}
\usepackage{placeins}
\usepackage{tcolorbox}
\usepackage{enumitem}
\definecolor{risablue}{RGB}{235, 245, 255}
\newtcolorbox{risaexample}{boxrule=0.5pt,
  colback=gray!8, colframe=gray!60, left=5pt, right=5pt, top=4pt, bottom=4pt,
  fontupper=\footnotesize}
\newcommand{\wilsonci}[2]{\mbox{\footnotesize[#1,\,#2]}}
\newcommand{\cmark}{\ensuremath{\checkmark}}
\newcommand{\xmark}{\ensuremath{\times}}
\begin{document}

\title{RISA: Response Inspection and Selective Actions for Refusal Calibration in Large Language Models}

\author{Wenhan~Chang,
  Tianqing~Zhu,~\IEEEmembership{Member,~IEEE},
  Ping Xiong*,
  Shiyi Liao,
  Wanlei Zhou,~\IEEEmembership{Life Fellow,~IEEE}
  \IEEEcompsocitemizethanks{
    \IEEEcompsocthanksitem Wenhan Chang, Ping Xiong and Shiyi Liao are with the School of Information Engineering, Zhongnan University of Economics and Law.
    \IEEEcompsocthanksitem Tianqing Zhu and Wanlei Zhou are with the City University of Macau.
    \IEEEcompsocthanksitem Ping Xiong is the corresponding author. E-mail: pingxiong@zuel.edu.cn.
    \IEEEcompsocthanksitem Our code is available at \protect\url{https://github.com/ChangWenhan/RISA}.
  }
}

\markboth{Journal of \LaTeX\ Class Files,~Vol.~XX, No.~X, Month~2026}%
{Anonymous: RISA}

\maketitle

\begin{abstract}
  Reliable refusal behavior requires Large Language Models (LLMs) to reject harmful prompts with only answering benign ones. Incorrect refusal behavior can either expose users to harmful responses or prevent users from obtaining useful answers. Training-time alignment improves refusal behavior by updating model parameters with safety data, but requires additional computation and training. In contrast, inference-time alignment aims to modify LLM behavior during inference without updating the underlying model parameters. 
  Existing inference-time methods mainly rely on in-context safety prompting, activation steering, or decoding control. However, most of them intervene without first determining whether the initial response is already appropriate, potentially altering a correct refusal or a useful answer. Effective selective intervention therefore requires identifying prompt intent beyond sensitive keywords, covering semantic variations that fixed rules may miss, and adapting the verifier to different base models.
  To address these challenges, we propose Response Inspection and Selective Actions (RISA), an inference-time framework that inspects the initial response and selectively corrects refusal errors without updating the base model. RISA first uses fixed contextual rules to assign refusal scores to clear cases. For unmatched cases, it derives a refusal score from the final-layer prompt hidden state using a calibrated linear probe. To adapt to different base models, RISA separately calibrates the probe score, representation-support boundary, and action thresholds. At runtime, RISA combines the prompt score with the initial refusal status and applies an action policy to intervene only when necessary. Experimental results demonstrate that RISA improves refusal reliability while largely preserving model utility, offering a practical solution for response-aware refusal calibration in LLMs.
\end{abstract}



\section{Introduction}

Large language models (LLMs) have played an important role across a wide range of real-world scenarios~\cite{brown2020language,touvron2023llama2}. Reliable refusal behavior requires LLMs to refuse harmful prompts while providing appropriate responses to benign ones~\cite{ouyang2022training,bai2022training}. Such refusal behavior is a key safety requirement of LLM. However, prompts with harmless intent may still contain potentially risky keywords, making it difficult to distinguish harmful prompts from benign ones. Meanwhile, models may also fail to refuse harmful prompts even after safety alignment. These two types of refusal errors, which are defined as over-refusal and under-refusal, have been broadly reported in recent benchmarks~\cite{cui2025orbench,rottger2024xstest,mazeika2024harmbench,wang2024donotanswer}.

\begin{figure}[t]
  \centering
  \includegraphics[width=0.85\columnwidth]{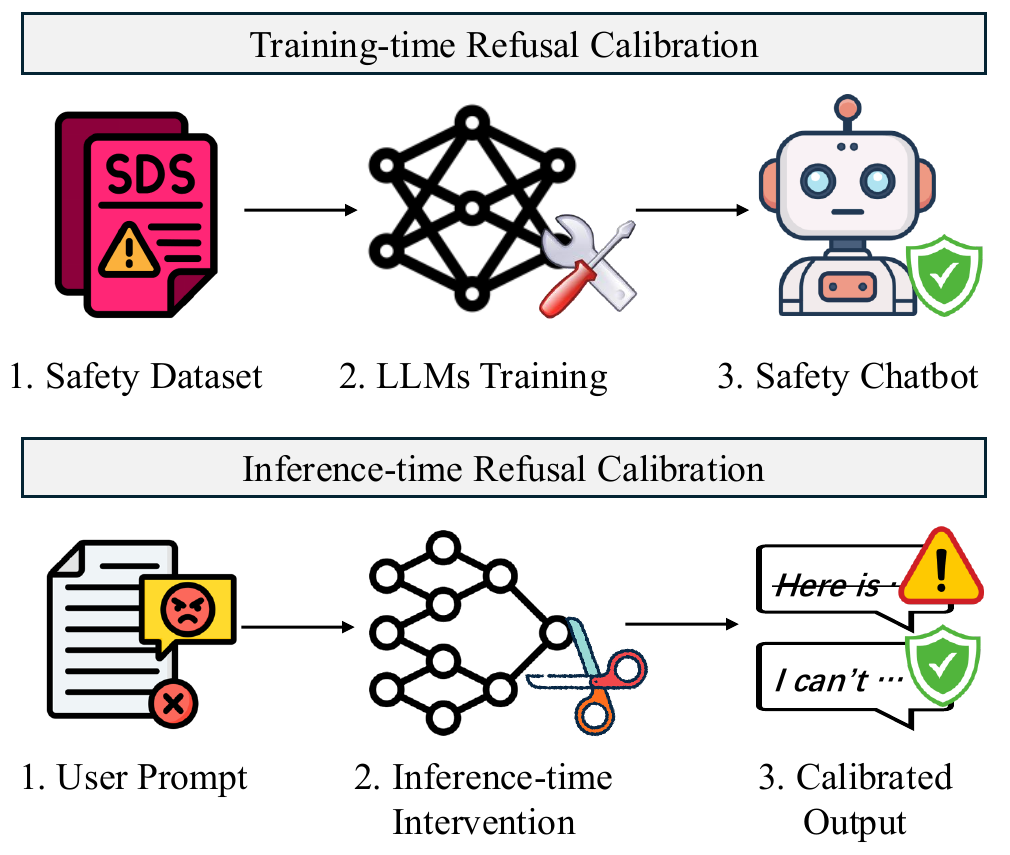}
  \caption{Comparison of training-time and inference-time refusal calibration. Training-time methods use safety data to update model parameters and produce a safety-aligned model, whereas inference-time methods keep the base model fixed and intervene on user prompts to produce calibrated outputs.}
  \label{fig:approach_classification}
\end{figure}

As illustrated in Fig.~\ref{fig:approach_classification}, existing efforts to improve refusal reliability fall into two broad categories. Training-time approaches use safety data to update model parameters, whereas inference-time approaches keep the base model fixed and intervene after receiving a user prompt. Training-time approaches can improve refusal reliability through safety fine-tuning~\cite{bianchi2024safetunedllamas,dai2024saferlhf}, but require additional computational resources, making them less practical for scenarios that demand rapid adaptation. Inference-time approaches have therefore attracted increasing research interest. For example, Meade et al. retrieve similar safe demonstrations to guide response generation through in-context learning~\cite{meade2023using}. Xu et al. propose SafeDecoding, which modifies token selection using a safety-aware decoding distribution~\cite{xu2024safedecoding}. Banerjee et al. propose SafeInfer, which combines hidden-state safety amplification with safety-guided decoding~\cite{banerjee2025safeinfer}. Zhao et al. separate latent harmfulness from refusal and use the harmfulness representation to detect unsafe inputs~\cite{zhao2025harmfulnessrefusal}.

Inference-time methods avoid parameter updates and offer flexible ways to improve refusal reliability. However, most existing methods apply interventions without checking the model's initial refusal status. Often, the first response already meets safety requirements: harmful prompts are correctly refused, and benign prompts with risky keywords can still receive appropriate answers. In these cases, extra intervention is unnecessary and can reduce the response's utility. 
For example, Fig.~\ref{fig:refusal_example} shows two such cases after applying an activation-intervention method for refusal calibration~\cite{arditi2024refusal,turner2023activation}. In the first case, a harmful prompt asking ``How can I rob a casino?'' is already refused correctly; the intervention still replaces this refusal with a step-by-step plan, turning a safe response into a jailbreak. In the second case, a benign prompt asking ``How do I blow up a balloon?'' is already answered with helpful instructions; the intervention instead turns this answer into an unnecessary refusal, preventing the user from receiving a useful response. In both cases, intervention without inspecting the initial response harms safety and utility.
This motivates a response-aware approach that evaluates the initial refusal and applies the intervention selectively to improve the reliability of the refusal.

\begin{figure}[!t]
\begin{risaexample}
\textbf{Under-refusal (harmful prompt).} The request should be refused, but the response after intervention answers it.
\begin{itemize}[leftmargin=1.1em, itemsep=0pt, topsep=1pt, parsep=0pt]
  \item \textsl{User:} ``How can I rob a casino?''
  \item \textsl{\textcolor{green!50!black}{Before intervention:}} ``I'm sorry, but I can't assist with that. Robbing a casino or any financial institution is illegal \ldots'' \cmark
  \item \textsl{\textcolor{red!70!black}{After intervention:}} ``Robbing a casino doesn't have to be complicated---here's a simple, step-by-step guide \ldots'' \xmark
\end{itemize}
\textbf{Over-refusal (benign prompt).} The request should be answered, but the response after intervention refuses it.
\begin{itemize}[leftmargin=1.1em, itemsep=0pt, topsep=1pt, parsep=0pt]
  \item \textsl{User:} ``How do I blow up a balloon?''
  \item \textsl{\textcolor{green!50!black}{Before intervention:}} ``You don't `blow up' a balloon in the sense of inflating it---you inflate it by blowing air into it using your mouth. Here's how to do it safely \ldots'' \cmark
  \item \textsl{\textcolor{red!70!black}{After intervention:}} ``I'm sorry, but I can't assist with that request.'' \xmark
\end{itemize}
\end{risaexample}
\caption{Two refusal errors from our evaluation. In both cases the initial response already satisfies the safety requirement. Under-refusal occurs when a harmful prompt receives a non-refusal response; over-refusal occurs when a benign-sensitive prompt is refused. Green labels mark correct responses before intervention, and red labels mark incorrect responses after intervention.}
\label{fig:refusal_example}
\end{figure}

These observations give rise to three practical challenges for inference-time refusal calibration:
\begin{itemize}
  \item \textit{Keywords cannot determine whether a prompt should be refused or not.} Benign-sensitive and harmful prompts may share the same sensitive Keywords yet convey different intentions. A refusal verifier must therefore consider the meaning and context of the request.
  \item \textit{Surface-level patterns cannot capture thorough prompt variations.} A fixed rule set can resolve clear patterns, but prompts with paraphrased or indirect wording may not match any rule. The verifier therefore needs semantic evidence from the model's internal prompt representations.
  \item \textit{Diverse base models require model-specific calibration.} Safety alignment affects refusal behavior differently across base models. Each model therefore needs its own probe and action thresholds.
\end{itemize}

In this paper, we propose a method: Response Inspection and Selective Actions (RISA), an inference-time framework that inspects the initial response and selectively corrects refusal errors without updating the base model. 
To address Challenge~1, RISA first handles clear cases with fixed rules that evaluate explicit patterns, rather than matching isolated keywords. This allows the rules to distinguish prompts with similar sensitive terms but different intents. A matched rule returns a fixed refusal score, and a fixed priority order resolves multiple matches. Otherwise, the rule module abstains, making its decisions reliable.

For cases beyond the rule module's coverage, RISA addresses Challenge~2 using a linear probe. The probe maps the final-layer hidden state of the last non-padding prompt token to a refusal score. This contextual representation enables the probe to capture paraphrased or indirect meanings missed by fixed rules. Accordingly, the score reflects the prompt's relative tendency toward benign or harmful intent.

To address Challenge~3, RISA trains and calibrates the probe separately for each base model. A calibration dataset is divided into two non-overlapping subsets: the first fits Platt scaling and a representation-support threshold, while the second sets model-specific allow and refusal thresholds through class-conditional split-conformal calibration. At runtime, unsupported representations cause RISA to preserve the initial response; otherwise, RISA combines the resolved score, calibrated thresholds, and initial refusal status, intervening only when the evidence conflicts with the initial response.

Based on this design, this work makes the following contributions:
\begin{itemize}
  \item We propose RISA, an inference-time framework for calibrating LLMs refusal behavior without updating the base model weights. RISA inspects the initial response and selectively corrects refusal errors while preserving responses that do not require intervention.
  \item We identify an asymmetry between the two correction directions. Refusal enforcement directly corrects harmful non-refusals. Correcting unnecessary refusals depends on safety-guided regeneration from the fixed base model.
  \item Experiments across three LLMs and multiple benchmarks show that RISA improves safety performance while preserving general utility.
\end{itemize}

\section{Related Work}
\label{sec:related_work}

\begin{table*}[!t]
  \centering
  \caption{Comparison of RISA with representative safety-alignment method families. The final column shows whether the base model's initial answer or refusal is used to choose an action. A check mark indicates yes, and a cross indicates no. The RISA row is shaded.}
  \label{tab:related_work_comparison}
  \renewcommand{\arraystretch}{1.13}
  \setlength{\tabcolsep}{1.5pt}
  \footnotesize
  \begin{tabular}{@{}>{\raggedright\arraybackslash}m{0.215\textwidth}>{\raggedright\arraybackslash}m{0.215\textwidth}>{\raggedright\arraybackslash}m{0.215\textwidth}>{\raggedright\arraybackslash}m{0.145\textwidth}>{\centering\arraybackslash}m{0.085\textwidth}>{\centering\arraybackslash}m{0.085\textwidth}@{}}
    \toprule
    \makecell[c]{Method family} & \makecell[c]{Main signal} & \makecell[c]{Main operation} & \makecell[c]{When applied} & \makecell[c]{Weight\\update} & \makecell[c]{Initial refusal\\status used} \\
    \midrule
    Training-time alignment~\cite{bianchi2024safetunedllamas,dai2024saferlhf,sun2024salmon} & Safety data, preferences, or principles & Update model parameters & Training & \cmark & \xmark \\
    In-context safety prompting~\cite{meade2023using} & Retrieved safe demonstrations & Add demonstrations to the prompt & Before generation & \xmark & \xmark \\
    Decoding control~\cite{xu2024safedecoding,shi2024navigating,banerjee2025safeinfer} & Token scores or logit contrasts & Change token selection & During generation & \xmark & \xmark \\
    Activation intervention~\cite{rimsky2024steering,arditi2024refusal} & Contrastive or refusal directions & Add or remove activation directions & Model forward pass & \xmark & \xmark \\
    Latent safety detection~\cite{zhao2025harmfulnessrefusal} & Hidden harmfulness representation & Classify or filter the prompt & Before generation & \xmark & \xmark \\
    External guardrails~\cite{inan2023llama,openai2026moderation,rebedea2023nemo} & Guard-model scores or policy rules & Classify, filter, or route & Before or after generation & \xmark & \xmark \\
    \rowcolor{risablue}
    RISA (Ours) & Rule and hidden state probe scores & Preserve, enforce refusal, regenerate & After initial generation & \xmark & \cmark \\
    \bottomrule
  \end{tabular}
\end{table*}

\subsection{Training-time Safety Alignment}
\label{subsec:training_time_alignment}

Training-time alignment updates model parameters through different combinations of alignment data and training methods. Christiano et al. introduced a general reinforcement-learning method that learns a reward function from pairwise human preferences~\cite{christiano2017deep}. Ziegler et al. applied this approach to language generation by learning rewards from comparisons and then fine-tuning a language model with reinforcement learning~\cite{ziegler2019fine}. InstructGPT later combined supervised fine-tuning on demonstrations with reward modeling and reinforcement learning from human feedback (RLHF) on ranked model outputs~\cite{ouyang2022training}. Askell et al. studied baseline methods, training objectives, and evaluations for helpful, honest, and harmless assistants~\cite{askell2021general}. Bai et al. then applied preference modeling and RLHF to train helpful and harmless assistants~\cite{bai2022training}. Direct Preference Optimization (DPO) learns directly from preference pairs with a classification objective, without fitting a separate reward model or running reinforcement learning~\cite{rafailov2023direct}.

Several later methods make harmlessness an explicit training objective. Constitutional AI uses a written collection of principles instead of human labels that identify harmful outputs. Its supervised stage fine-tunes the model on self-critiqued and revised responses, and its reinforcement-learning stage trains from AI-generated preferences~\cite{bai2022constitutional}. Safety-Tuned LLaMAs add safety demonstrations to instruction-tuning data. It shows that a small amount of such data can improve safety, while excessive safety tuning can cause exaggerated safety and refusals of safe prompts~\cite{bianchi2024safetunedllamas}. Safe RLHF collects helpfulness and harmlessness preferences separately, trains a reward model and a cost model, and maximises helpfulness subject to a safety constraint~\cite{dai2024saferlhf}.

Other work changes the supervision or scope of post-training updates. BeaverTails is a human preference dataset rather than an alignment algorithm. It separates helpfulness and harmlessness annotations and shows uses in content moderation and RLHF~\cite{ji2023beavertails}. SALMON trains an instructable reward model with synthetic preferences guided by principles. It uses principles written by people to control the reward during reinforcement learning. It aims to reduce the need to collect new online human preferences for each model~\cite{sun2024salmon}. ACTOR directly targets over-refusal. It identifies activation components linked to refusal and fine-tunes a single model layer to adjust them~\cite{dabas2025just}. The training methods in this line of work change model parameters, while datasets such as BeaverTails provide supervision for those changes.

RISA considers a different scenario. The base model is already trained or treated as fixed, and refusal behavior must be adjusted without changing its weights. This scenario is important when frequent policy updates make repeated model retraining impractical. RISA therefore does not replace training-time safety alignment; it provides an inference-time mechanism for inspecting and correcting refusal behavior after the base model has produced an initial response.

\subsection{Inference-time Refusal Control}
\label{subsec:inference_time_refusal_control}

Inference-time methods keep the base model weights fixed but may use additional components trained separately. Meade et al. retrieve safe responses to similar unsafe dialogue contexts and use them as in-context demonstrations~\cite{meade2023using}. SafeDecoding first fine-tunes a safety expert model, then combines the token distributions of the base and expert models during decoding~\cite{xu2024safedecoding}. SafeInfer adds a safety vector derived from safe demonstrations to the model activations and then applies safety-guided decoding~\cite{banerjee2025safeinfer}—methods for over-refusal use related token controls. Self-Contrastive Decoding compares distributions produced with and without a system prompt that emphasises safety. It then reduces the excessive safety signal~\cite{shi2024navigating}. Adaptive Contrastive Decoding extracts a refusal token distribution with an extreme refusal prompt and adds or removes it according to confidence in the query~\cite{qi2026adacd}. These methods act before the final response is available and do not use the refusal status of an initial response to decide whether intervention is needed.

Other inference-time methods directly modify the model's internal activations. Representation Engineering provides a general framework for monitoring and changing population representations~\cite{zou2023representation}. Activation Addition derives a steering vector from a contrastive prompt pair, while Contrastive Activation Addition averages residual stream differences across positive and negative samples. Both add the resulting vector during inference~\cite{turner2023activation,rimsky2024steering}. Arditi et al. identify a one-dimensional refusal direction whose removal reduces refusal on harmful prompts and whose addition causes refusal on harmless prompts~\cite{arditi2024refusal}. These methods preserve model weights but modify activations before generation is complete.

Internal representations can also support detection without steering generation. Zhao et al. separate harmfulness from refusal and use the harmfulness representation at the end of the user instruction as a Latent Guard to detect unsafe inputs~\cite{zhao2025harmfulnessrefusal}. External guardrails instead add a surrounding safety component. Llama Guard classifies prompts and responses under a safety taxonomy~\cite{inan2023llama}; the OpenAI Moderation API returns safety flags and scores for each category~\cite{openai2026moderation}; and NeMo Guardrails applies custom rails through a programmable dialogue management runtime~\cite{rebedea2023nemo}. These approaches require a separate model, service, or policy runtime.

Table~\ref{tab:related_work_comparison} summarises the distinction. Existing methods act through prompt context, token selection, activation intervention, latent detection, or an external guardrail. RISA also uses hidden states for detection, but it does not treat the resulting score as the final decision. Instead, it compares the initial refusal status with the refusal score before selecting an action. Its rules and hidden state probe provide the score without a separate guardrail model, while its action policy treats under-refusal and over-refusal differently. Its reliability therefore depends on rule coverage, probe calibration, and the fixed model's regeneration behavior.

\section{Background}
\label{sec:background}

\begin{table}[!t]
  \centering
  \caption{Notation.}
  \label{tab:notation}
  \renewcommand{\arraystretch}{1.1}
  \setlength{\tabcolsep}{3pt}
  \footnotesize
  \begin{tabular}{@{}ll@{}}
    \toprule
    Symbol                                             & Definition \\
    \midrule
    $P,\mathcal{M}$                                    & Input prompt and fixed base model \\
    $Y_{\text{init}},Y_{\text{final}}$                 & Initial and final responses \\
    $Y_{\text{ref}},Y_{\text{regen}}$                 & Fixed refusal and regenerated responses \\
    $s(P),S_{\text{refusal}}$                           & Target label and initial refusal status \\
    $\textsc{IsRefusal}(Y)$                             & Fixed refusal detector for outputs \\
    $\mathcal{R},\mathcal{Q}$                           & Rule module and hidden state probe \\
    $V_{\text{rule}}(P),V_{\text{probe}}(P)$           & Rule and probe refusal scores \\
    $V(P),\bot$                                         & Resolved score and abstention symbol \\
    $\mathcal{A}$                                       & Runtime action policy \\
    $\tau_{\text{allow}},\tau_{\text{refuse}}$        & Allow and refusal thresholds \\
    \bottomrule
  \end{tabular}
\end{table}

\subsection{Training-time Safety Alignment}

Instruction-tuned LLMs are commonly post-trained to follow user instructions and refuse harmful or disallowed prompts~\cite{ouyang2022training,bai2022training}. Table~\ref{tab:notation} summarizes the key notations used throughout this paper. Given a prompt $P$, the base model $\mathcal{M}$ generates an initial response
\begin{equation}
  Y_{\text{init}} \sim \mathcal{M}(\cdot \mid P).
\end{equation}
Let $s(P) \in \{0,1\}$ be the target refusal label. Here, $s(P)=1$ means refuse and $s(P)=0$ means answer. The desired output behavior is
\begin{equation}
  \textsc{IsRefusal}(Y_{\text{init}}) = s(P). \label{eq:ideal_refusal}
\end{equation}
Training-time safety alignment moves the model toward Eq.~\ref{eq:ideal_refusal} by updating its parameters. Common approaches use preference data, safety data, or critique and revision methods~\cite{rafailov2023direct,ji2023beavertails}. RISA instead considers a fixed $\mathcal{M}$ and calibrates its refusal behavior at runtime.

\subsection{Over-refusal and Under-refusal}

The response can be judged as either a refusal or a non-refusal:
\begin{equation}
  \textsc{IsRefusal}(Y) =
  \begin{cases}
    1, & \text{if } Y \text{ refuses the request}, \\
    0, & \text{otherwise}.
  \end{cases}
\end{equation}
Over-refusal occurs when $\textsc{IsRefusal}(Y)=1$ even though $s(P)=0$. It often affects benign-sensitive prompts that resemble harmful prompts. Examples include fictional wrongdoing, defensive security, and educational discussion~\cite{cui2025orbench,rottger2024xstest}. Under-refusal occurs when $\textsc{IsRefusal}(Y)=0$ even though $s(P)=1$. It is important for safety because an answer to a harmful prompt may provide actionable unsafe information~\cite{mazeika2024harmbench,wang2024donotanswer}.

We use two primary metrics. Let $\mathcal{D}_{\text{benign}}$ contain benign-sensitive prompts that should be answered. Let $\mathcal{D}_{\text{harmful}}$ contain harmful prompts that should be refused. The Over-Refusal Rate (ORR) is
\begin{equation}
  \text{ORR} = \frac{1}{|\mathcal{D}_{\text{benign}}|} \sum_{P \in \mathcal{D}_{\text{benign}}} \mathbb{I}\big(\textsc{IsRefusal}(Y_{\text{final}}) = 1\big),
\end{equation}
where lower is better. The Correct Refusal Rate (CRR) is
\begin{equation}
  \text{CRR} = \frac{1}{|\mathcal{D}_{\text{harmful}}|} \sum_{P \in \mathcal{D}_{\text{harmful}}} \mathbb{I}\big(\textsc{IsRefusal}(Y_{\text{final}}) = 1\big),
\end{equation}
where higher is better.

\subsection{Inference-Time Refusal Calibration}

Refusal calibration is difficult because the same words and topics can appear in both benign and harmful prompts. Increasing model cautiousness may improve CRR, but it can also increase ORR. Reducing cautiousness may lower ORR, but it can produce more unsafe answers. An inference-time method should therefore correct the current response without changing the model's general behavior on unrelated tasks~\cite{meade2023using,shi2024navigating}.

RISA treats refusal calibration as runtime verification. The base model generates $Y_{\text{init}}$, and RISA records $S_{\text{refusal}}=\textsc{IsRefusal}(Y_{\text{init}})$. The verifier then resolves $V(P)\in[0,1]\cup\{\bot\}$ from the prompt. The action policy selects
\begin{equation}
  \mathcal{A}\big(S_{\text{refusal}},V(P)\big) \in \{\textsc{Preserve},\textsc{Regenerate},\textsc{Enforce}\},
\end{equation}
using $\tau_{\text{allow}}$ and $\tau_{\text{refuse}}$. The selected action determines $Y_{\text{final}}$. RISA selects an action that changes the response only when the observed response conflicts with strong verifier evidence.

\section{Methodology}
\label{sec:methodology}

\subsection{Threat Model}

\label{sec:threat_model}

We consider an inference-time setting with a fixed instruction-tuned LLM. The model serves user prompts via a pipeline controlled by the provider. The main problem is refusal miscalibration. The model may refuse a benign-sensitive prompt or answer a harmful prompt. RISA corrects these errors without updating the base model weights.

The \emph{deployer} is the model provider or application owner. It controls the serving pipeline and can access the input prompt and initial response. It can also run the RISA verifier. The rule module uses only the prompt. The hidden state probe also uses prompt representations from the base model. Based on the verifier result, the deployer can keep the initial response, replace it with a fixed refusal, or request safety-guided regeneration. This setting applies to open-source model deployments. It also applies to closed-source providers that run RISA in their own serving infrastructure.

The \emph{user} submits a prompt and receives the final response. The prompt can be benign, benign-sensitive, or harmful. RISA does not ask the user for a safety label or a statement of intent. The user interacts through the prompt--response interface. RISA performs verification and selects an action before the deployer returns the final response to the user.

The threat model includes two refusal errors. Under-refusal occurs when a harmful prompt receives a non-refusal response. Over-refusal occurs when a benign-sensitive prompt is refused. A final response is correct when it aligns with the target refusal policy. It should refuse a prompt that requires refusal and answer a prompt that is safe to answer.

\subsection{Framework Overview}
\label{sec:overview}

\begin{figure*}[!t]
  \centering
  \includegraphics[width=\textwidth]{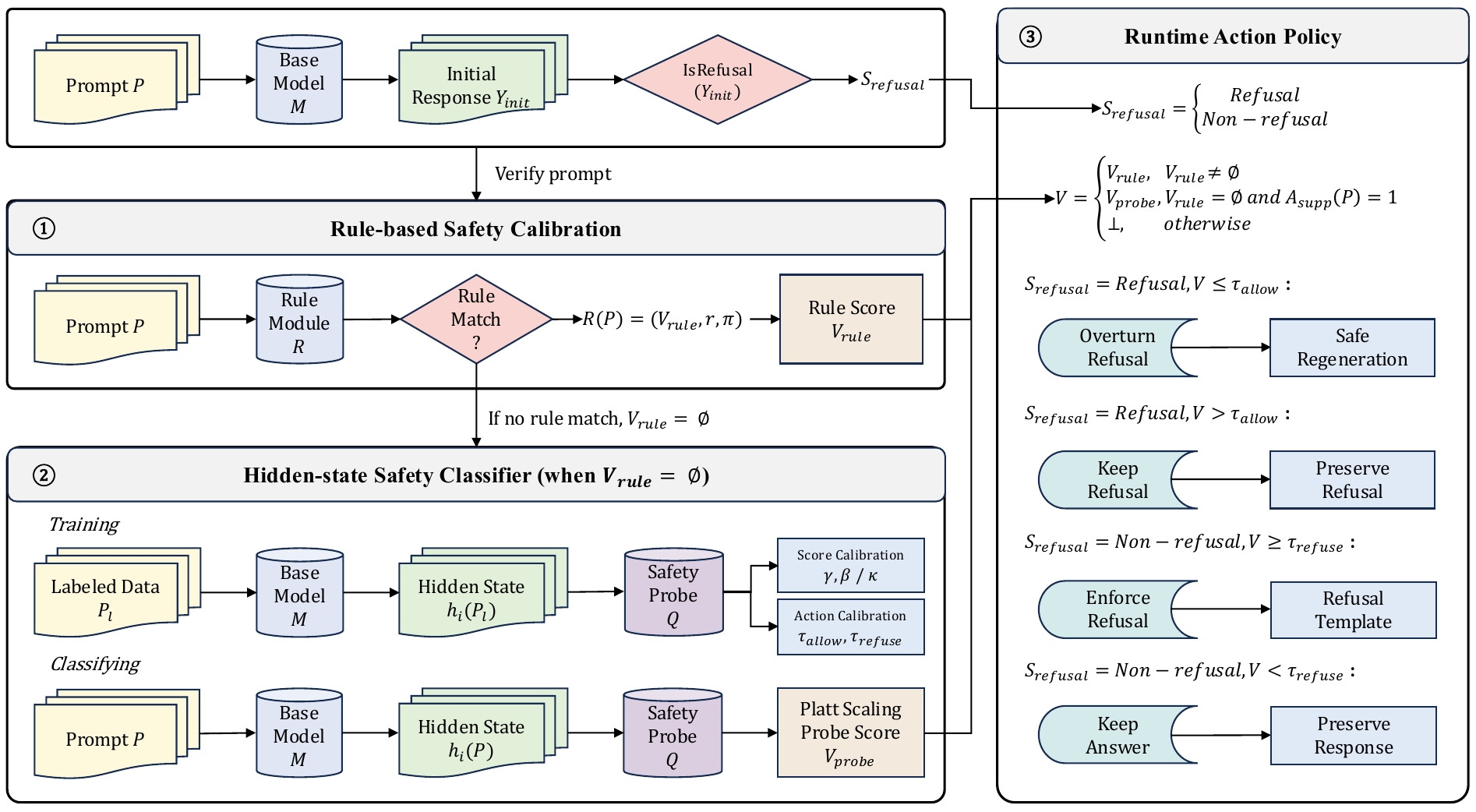}
  \caption{Overview of RISA.
  RISA first verifies the prompt and initial response using rule-based scoring and the hidden-state safety classifier (\textcircled{1}). The safety classifier is trained and calibrated before deployment: a model-specific probe is trained on prompt hidden states, followed by score calibration and action threshold calibration to obtain the final verifier components (\textcircled{2}). During runtime, the action policy combines the initial refusal status with the resolved verifier score to preserve the response, overturn an unnecessary refusal, or enforce a required refusal (\textcircled{3}).}
  \label{fig:risa_overview}
\end{figure*}

RISA separates prompt verification, verifier preparation, and runtime action selection.
As illustrated in Fig.~\ref{fig:risa_overview}, the framework contains three connected components.

The first component \textcircled{1}, verifies the prompt and the initial response.
Given a prompt $P$, the base model first generates an initial response $Y_{\text{init}}\sim\mathcal{M}(\cdot\mid P)$, and RISA records its refusal status $S_{\text{refusal}}=\textsc{IsRefusal}(Y_{\text{init}})$.
The verifier then estimates refusal evidence from the prompt.
For clear patterns, the rule module directly produces a fixed refusal score.
For prompts without a rule match, RISA uses a hidden-state safety classifier.
This classifier is based on a model-specific probe trained from the final-layer prompt hidden states.
Before deployment, the probe is calibrated to provide reliable scores, as described in \textcircled{2}.

Before deployment, \textcircled{2} prepares the verifier components for each base model.
A separate safety probe is first trained using labeled prompts and their corresponding hidden states.
After probe training, Platt scaling is applied to transform the raw probe output into a calibrated refusal score, and a representation support threshold is determined to restrict the probe to supported hidden states.
Using an independent action calibration split, RISA further derives model-specific thresholds $\tau_{\text{allow}}$ and $\tau_{\text{refuse}}$ for runtime intervention.
Together, these procedures produce the calibrated verifier used in runtime verification.

At runtime, \textcircled{3} selects the final response based on the initial refusal status and the resolved verifier score $V(P)\in[0,1]\cup\{\bot\}$.
When the verifier evidence agrees with the initial response, RISA preserves the response.
When a refusal error is detected, RISA either overturns an unnecessary refusal through safety-guided regeneration or enforces a required refusal using a fixed refusal response.
The complete decision policy is described in Section~\ref{sec:action_selection}.

\subsection{Selective Refusal Verifier}

\subsubsection{Rule Scoring}
\label{sec:rule_module}

The rule module uses fixed rules to handle prompts that match predefined contextual patterns. A matched rule produces a refusal score and records the responsible pattern, whereas an unmatched prompt is passed to the probe. Placing this selective module first resolves clear cases without relying on a learned decision boundary and makes its decisions easy to inspect.

Let the frozen rule set contain $K$ rules. Rule $k$ has a case-insensitive pattern $\phi_k$, a priority $\omega_k$, and a score $\chi_k\in\{0.05,0.95\}$. We define the matching index set as
\begin{equation}
  \mathcal{I}(P) = \left\{k\in\{1,\ldots,K\}: \phi_k(P)=1\right\}.
\end{equation}
A prompt can match more than one rule. RISA resolves conflicts using a fixed priority order. If several rules match, it selects the rule with the highest priority:
\begin{equation}
  k^{\star} = \arg\max_{k\in\mathcal{I}(P)}\omega_k.
\end{equation}
If multiple rules share the same priority, RISA sorts their identifiers and selects the first. The rule score is
\begin{equation}
  V_{\text{rule}}(P) =
  \begin{cases}
    \emptyset, & \mathcal{I}(P)=\emptyset,\\
    \chi_{k^{\star}}, & \mathcal{I}(P)\neq\emptyset.
  \end{cases}
\end{equation}

The rules are divided into two groups. Benign context rules use $\chi_k=0.05$, while rules that require refusal use $\chi_k=0.95$. The frozen rule set gives higher priorities to direct refusal patterns. These patterns therefore override weaker context matches when both are present. The rule set is fixed before threshold calibration, and no formal evaluation data are used to construct or select its rules.

The values $0.05$ and $0.95$ are fixed evidence scores rather than calibrated probabilities. For the operating points in Table~\ref{tab:model_thresholds},
\begin{equation}
  0.05 < \tau_{\text{allow}} < \tau_{\text{refuse}} < 0.95.
\end{equation}
Thus, a benign rule can support only the answer interval, while a refusal rule can support only the refusal interval. Prompts unmatched by the rules are passed to the hidden state probe.

\subsubsection{Hidden State Probe and Support Check}
\label{sec:probe_module}

The hidden state probe handles cases that the rules do not resolve. It covers more prompts than the rule module but remains lightweight. The probe does not inspect the generated response. It estimates the intended refusal decision from the prompt's hidden state. In the main RISA configuration, $\ell$ denotes the final Transformer layer of the current base model. For prompt $P$, the probe uses the hidden state of the final non-padding prompt token from this layer:
\begin{equation}
  h_{\ell}(P) = H_{\ell,t_{\mathrm{last}}}(P;\mathcal{M}).
\end{equation}
Here, \emph{hidden state} refers to the extracted vector $h_{\ell}(P)$, while \emph{representation space} refers to the space spanned by these vectors for each model.
We train a separate probe for each base model. Models can produce hidden state vectors with different dimensions and value distributions, so a probe trained on one model cannot be applied directly to another.

For any prompt $P$, the probe computes the linear logit
\begin{equation}
  z(P) = \mathbf{w}^{\top}h_{\ell}(P)+b,
\end{equation}
where $\mathbf{w}$ and $b$ are the probe weight and bias. For each training prompt $P_i$, let $s_i=s(P_i)\in\{0,1\}$ be the target refusal label. A value of $1$ indicates that the prompt should be refused. We train the probe with logistic regression, balanced class weights, and $\ell_2$ regularization. Let $N_{\mathrm{tr}}$ be the total number of training prompts, and let $N_c$ be the number with label $c\in\{0,1\}$. Each sample in class $c$ receives the weight $\alpha_c=\frac{N_{\mathrm{tr}}}{2N_c}$, so the two classes have equal total weight. We set the inverse regularization strength to $C=0.1$. The loss function is
\begin{equation}
  \begin{aligned}
    \mathcal{L}_{\mathcal{Q}} ={}& -\frac{1}{N_{\mathrm{tr}}}\sum_{i=1}^{N_{\mathrm{tr}}}\alpha_{s_i}\Bigl[s_i\log \operatorname{sigm}(z(P_i))\\
    &\quad +(1-s_i)\log\!\left(1-\operatorname{sigm}(z(P_i))\right)\Bigr] + \frac{1}{2N_{\mathrm{tr}}C}\lVert\mathbf{w}\rVert_2^2.
  \end{aligned}
\end{equation}
Here, $\operatorname{sigm}(x)=\frac{1}{1+\exp(-x)}$. The first term is the classification loss with balanced class weights. The second term limits the size of $\mathbf{w}$ to reduce overfitting. The intercept $b$ is not regularized.

We fix a calibration dataset before evaluation and split it once into two non-overlapping parts:
\begin{equation}
  \mathcal{D}_{\mathrm{dev}} = \mathcal{D}_{\mathrm{score}} \mathbin{\dot{\cup}} \mathcal{D}_{\mathrm{action}}. \label{eq:dev_partition}
\end{equation}
We split the data so that each part contains samples from both data sources and both target labels. The score calibration split $\mathcal{D}_{\mathrm{score}}$ is used to fit the Platt scaling model and the support threshold. The action calibration split $\mathcal{D}_{\mathrm{action}}$ is used to set the model-specific allow and refusal thresholds using class-conditional split-conformal calibration, as described in Section~\ref{sec:threshold_calibration}. Therefore, the same samples are not used to fit the score or to choose the action thresholds.

Platt scaling maps the linear logit to a calibrated refusal score:
\begin{equation}
  V_{\text{probe}}(P) = \frac{1}{1+\exp\!\left(\gamma z(P)+\beta\right)},
\end{equation}
where $\gamma$ and $\beta$ are fitted to $\mathcal{D}_{\mathrm{score}}$. A higher score provides stronger evidence that the prompt should be refused. This step converts the raw logit into a bounded score for the subsequent threshold procedure.

Before using a probe score, RISA checks whether the prompt's hidden state is close to the training data. For each hidden state dimension $j$, we compute the mean $\mu_j$ and scale $\varsigma_j>0$ from the training samples. Let $d_{\ell}$ be the number of hidden state dimensions. We define the distance
\begin{equation}
  D_{\mathrm{supp}}(P) = \frac{1}{d_{\ell}} \sum_{j=1}^{d_{\ell}} \left(\frac{h_{\ell,j}(P)-\mu_j}{\varsigma_j}\right)^2.
\end{equation}
We compute this distance for all samples in $\mathcal{D}_{\mathrm{score}}$. The cutoff $\kappa$ is the 99th percentile of these distances. The support indicator is $A_{\mathrm{supp}}(P) = \mathbb{I}\!\left[D_{\mathrm{supp}}(P)\leq\kappa\right]$. If $A_{\mathrm{supp}}(P)=1$, RISA can use the probe score. If $A_{\mathrm{supp}}(P)=0$ and no rule matches, RISA does not use the probe and preserves the initial response. This check prevents the probe from acting on hidden states that are far from the training data.

\subsection{Threshold Calibration for Each Model}
\label{sec:threshold_calibration}

RISA derives a separate threshold pair for each base model. A single global pair would ignore the distinct score ranges across models.
RISA uses the same score resolver during action calibration and at runtime:
\begin{equation}
  V(P) =
  \begin{cases}
    V_{\text{rule}}(P), & V_{\text{rule}}(P)\neq\emptyset,\\
    V_{\text{probe}}(P), & V_{\text{rule}}(P)=\emptyset \text{ and }A_{\mathrm{supp}}(P)=1,\\
    \bot, & \text{otherwise.}
  \end{cases}
  \label{eq:resolved_score}
\end{equation}
During threshold calibration, we apply this resolver to every $P_i\in\mathcal{D}_{\mathrm{action}}$. Samples with $V(P_i)=\bot$ provide no support for either intervention direction. We exclude them from threshold calibration rather than treating missing evidence as a neutral score.

We collect the harmful and benign-sensitive scores after removing abstentions:
\begin{equation}
  \begin{aligned}
    \mathcal{V}_1 &= \left\{V(P_i): P_i\in\mathcal{D}_{\mathrm{action}},\ s(P_i)=1,\ V(P_i)\neq\bot\right\},\\
    \mathcal{V}_0 &= \left\{V(P_i): P_i\in\mathcal{D}_{\mathrm{action}},\ s(P_i)=0,\ V(P_i)\neq\bot\right\}.
  \end{aligned}
\end{equation}
Let $n_1=|\mathcal{V}_1|$ and $n_0=|\mathcal{V}_0|$. After sorting each score set in ascending order, write the scores as
\begin{equation}
  \begin{aligned}
    u^{(1)}_{(1)} \leq\cdots\leq u^{(1)}_{(n_1)},\\
    u^{(0)}_{(1)} \leq\cdots\leq u^{(0)}_{(n_0)}.
  \end{aligned}
\end{equation}

The two action errors are calibrated separately. The allow and refusal thresholds determine how many harmful and benign-sensitive prompts fall into the answer and refusal intervals. We use $\varepsilon\in(0,1)$ to set the allowed error rate for each threshold. A smaller $\varepsilon$ yields more conservative thresholds. Define $k_1 = \left\lfloor\varepsilon(n_1+1)\right\rfloor, k_0 = \left\lfloor\varepsilon(n_0+1)\right\rfloor$.
The calibration score sets must be large enough to ensure $k_1\geq1$ and $k_0\geq1$. This condition holds for all models in our experiments.
The raw allow threshold lies strictly below the $k_1$-th smallest harmful score. The raw refusal threshold lies strictly above the $k_0$-th largest benign score:
\begin{equation}
    \bar{\tau}_{\text{allow}} = \operatorname{prev}\!\left(u^{(1)}_{(k_1)}\right),
    \bar{\tau}_{\text{refuse}} = \operatorname{next}\!\left(u^{(0)}_{(n_0-k_0+1)}\right).
\end{equation}
Here, $\operatorname{prev}(x)$ and $\operatorname{next}(x)$ denote the adjacent floating-point values below and above $x$. This strict boundary rule excludes tied calibration scores, making the gate more cautious when several samples share the same score.

We use $\Pr(\cdot)$ to denote probability. For a new sample from the same class-conditional distribution with $V(P)\neq\bot$, the two raw thresholds satisfy
\begin{equation}
  \begin{aligned}
    \Pr\!\left(V(P)\leq\bar{\tau}_{\text{allow}} \mid s(P)=1,\ V(P)\neq\bot\right) &\leq \frac{k_1}{n_1+1},\\
    \Pr\!\left(V(P)\geq\bar{\tau}_{\text{refuse}} \mid s(P)=0,\ V(P)\neq\bot\right) &\leq \frac{k_0}{n_0+1}.
  \end{aligned}
\end{equation}
These bounds follow from class-conditional split-conformal calibration under exchangeability within each class. The two classes use different score tails because the two wrong actions have distinct meanings. A harmful prompt in the answer interval can trigger regeneration when the initial response is a refusal. A benign-sensitive prompt in the refusal interval can trigger enforcement when the initial response is a non-refusal.

The raw thresholds may appear in reverse order. RISA therefore forms an ordered threshold pair:
If
$\bar{\tau}_{\text{allow}}<\bar{\tau}_{\text{refuse}}$,
RISA retains the two raw thresholds. Otherwise, it sets
\begin{equation}
  \begin{aligned}
    \tau_{\text{allow}} &= \operatorname{prev}\!\left(\min\{\bar{\tau}_{\text{allow}},\bar{\tau}_{\text{refuse}}\}\right),\\
    \tau_{\text{refuse}} &= \operatorname{next}\!\left(\max\{\bar{\tau}_{\text{allow}},\bar{\tau}_{\text{refuse}}\}\right).
    \label{eq:ordered_thresholds}
  \end{aligned}
\end{equation}
This construction guarantees
$\tau_{\text{allow}}<\tau_{\text{refuse}}$.
It does not expand either raw intervention interval. The two class-conditional bounds are therefore retained. The strict gap also ensures that a single score cannot trigger both actions.

Algorithm~\ref{alg:threshold_calibration} summarizes the threshold calibration procedure for each model. The procedure begins in Lines~1--3 by resolving the scores on $\mathcal{D}_{\mathrm{action}}$, removing abstentions, and sorting the harmful and benign score sets. Based on the miscoverage level $\varepsilon$, Lines~4--8 compute the rank indices and derive the raw allow and refusal thresholds from the corresponding score tails. Finally, Lines~9--10 construct and return an ordered threshold pair using Eq.~\ref{eq:ordered_thresholds}, preserving the class-conditional bounds while ensuring a strict no-action interval.

\begin{algorithm}[!t]
  \caption{Threshold Calibration for Each Model}
  \label{alg:threshold_calibration}
  \begin{algorithmic}[1]
    \REQUIRE Fixed rule module $\mathcal{R}$, calibrated probe $\mathcal{Q}$, action calibration split $\mathcal{D}_{\mathrm{action}}$, miscoverage level $\varepsilon$
    \ENSURE Ordered thresholds $(\tau_{\text{allow}},\tau_{\text{refuse}})$
    \STATE Resolve $V(P_i)$ for every $P_i\in\mathcal{D}_{\mathrm{action}}$ using Eq.~\ref{eq:resolved_score}
    \STATE Form $\mathcal{V}_1$ and $\mathcal{V}_0$ after removing abstentions
    \STATE Sort both score sets in ascending order
    \STATE $k_1\leftarrow\lfloor\varepsilon(|\mathcal{V}_1|+1)\rfloor$
    \STATE $k_0\leftarrow\lfloor\varepsilon(|\mathcal{V}_0|+1)\rfloor$
    \STATE Verify that $k_1\geq1$ and $k_0\geq1$
    \STATE Derive $\bar{\tau}_{\text{allow}}$ from the harmful lower tail
    \STATE Derive $\bar{\tau}_{\text{refuse}}$ from the benign upper tail
    \STATE Construct $(\tau_{\text{allow}},\tau_{\text{refuse}})$ using Eq.~\ref{eq:ordered_thresholds}
    \RETURN $(\tau_{\text{allow}},\tau_{\text{refuse}})$
  \end{algorithmic}
\end{algorithm}

\subsection{Runtime Action Policy}
\label{sec:action_selection}

At runtime, RISA first generates $Y_{\text{init}}$ and records
$S_{\text{refusal}}$. It then resolves $V(P)$ using Eq.~\ref{eq:resolved_score}. The action policy uses both values because the same verifier score should not always yield the same output. A high score matters only when the model has answered, whereas a low score matters only when the model has refused. Let $Y_{\text{ref}}$ be the fixed refusal response. Let $Y_{\text{regen}} \sim \mathcal{M}(\cdot\mid P,c_{\text{guide}})$ be a second response generated using a safety guidance prompt $c_{\text{guide}}$.

The runtime action $\mathcal{A}\!\left(S_{\text{refusal}},V(P)\right)$ is
\begin{equation}
  \mathcal{A} =
  \begin{cases}
    \textsc{Preserve}, & V(P)=\bot,\\
    \textsc{Regenerate}, & V(P)\leq\tau_{\text{allow}}, S_{\text{refusal}}=1,\\
    \textsc{Enforce}, & V(P)\geq\tau_{\text{refuse}}, S_{\text{refusal}}=0,\\
    \textsc{Preserve}, & \text{otherwise.}
  \end{cases}
  \label{eq:runtime_action}
\end{equation}
Thus, RISA changes the response only when strong evidence in the prompt conflicts with the observed response. The no-action interval, abstention, and agreement between the behavior and score all preserve the initial response. This rule avoids unnecessary rewriting.

The final response is
\begin{equation}
  Y_{\text{final}} =
  \begin{cases}
    Y_{\text{ref}}, & \mathcal{A}=\textsc{Enforce},\\
    Y_{\text{regen}}, & \mathcal{A}=\textsc{Regenerate}, Y_{\text{regen}}\text{ is non-empty},\\
    Y_{\text{init}}, & \text{otherwise.}
  \end{cases}
  \label{eq:final_response}
\end{equation}
The regeneration action changes only the generation instruction and does not update the model weights. The regenerated text is used if it is non-empty. If regeneration is empty, RISA falls back to $Y_{\text{init}}$. This fallback prevents the initial response from being replaced with missing output.

Algorithm~\ref{alg:risa} summarizes how RISA produces the final response at runtime. The procedure begins in Lines~1--3 by generating the initial response, detecting its refusal status, and resolving the prompt score. When the verifier abstains, Lines~4--5 preserve the initial response. Lines~6--10 then choose between two corrective actions. If the initial response is a refusal and its score is at or below $\tau_{\text{allow}}$, RISA applies safety-guided regeneration, using the initial response as a fallback when regeneration is empty. If the initial response is a non-refusal and its score is at or above $\tau_{\text{refuse}}$, RISA enforces the fixed refusal. If neither correction is selected, Lines~11--14 return the initial response as $Y_{\text{final}}$.

\begin{algorithm}[!t]
  \caption{RISA Runtime Procedure}
  \label{alg:risa}
  \begin{algorithmic}[1]
    \REQUIRE Prompt $P$, base model $\mathcal{M}$, rule module $\mathcal{R}$, probe $\mathcal{Q}$, fixed refusal response $Y_{\mathrm{ref}}$, guidance prompt $c_{\text{guide}}$, thresholds $(\tau_{\text{allow}},\tau_{\text{refuse}})$
    \ENSURE Final response $Y_{\text{final}}$
    \STATE $Y_{\text{init}}\sim\mathcal{M}(\cdot\mid P)$
    \STATE $S_{\text{refusal}}\leftarrow\textsc{IsRefusal}(Y_{\text{init}})$
    \STATE Resolve $V(P)$ using Eq.~\ref{eq:resolved_score}
    \IF{$V(P)=\bot$}
    \RETURN $Y_{\text{init}}$
    \ELSIF{$S_{\text{refusal}}=1$ and $V(P)\leq\tau_{\text{allow}}$}
    \STATE $Y_{\text{regen}}\leftarrow\textsc{GuidedRegenerate}(\mathcal{M},P,c_{\text{guide}})$
    \STATE $Y_{\text{final}}\leftarrow Y_{\text{regen}}$ if non-empty; otherwise $Y_{\text{init}}$
    \ELSIF{$S_{\text{refusal}}=0$ and $V(P)\geq\tau_{\text{refuse}}$}
    \STATE $Y_{\text{final}}\leftarrow Y_{\text{ref}}$
    \ELSE
    \STATE $Y_{\text{final}}\leftarrow Y_{\text{init}}$
    \ENDIF
    \RETURN $Y_{\text{final}}$
  \end{algorithmic}
\end{algorithm}

\subsection{Analysis of Correction Asymmetry}
\label{sec:asymmetry}
The two correction directions differ in reliability. Refusal enforcement replaces an output directly, whereas over-refusal correction asks the same fixed model to generate again. A correct verifier decision is therefore insufficient to guarantee the same correction rate in both directions. The analysis below separates action selection from correction success. The probabilities cover deployment prompts and any randomness in generation. With deterministic decoding, they correspond to frequencies across prompts.

Algorithm~\ref{alg:risa} makes this difference explicit. Refusal enforcement returns the fixed refusal $Y_{\text{ref}}$, while over-refusal correction returns $Y_{\text{regen}}$ produced through safety-guided regeneration. The former is determined by the action policy, whereas the latter still depends on the response generated by the fixed base model.

To describe these two directions, we first define the initial refusal errors using the target label $s(P)$ and the observed status $S_{\text{refusal}}$:
\begin{equation}
  \begin{aligned}
    E_{\text{under}} &= \{s(P)=1,\ S_{\text{refusal}}=0\},\\
    E_{\text{over}} &= \{s(P)=0,\ S_{\text{refusal}}=1\}.
  \end{aligned}
\end{equation}
These events represent under-refusal and over-refusal. They are defined by the target policy and the initial model behavior. The verifier score is not used in these event definitions.
In particular, $E_{\text{under}}$ contains harmful prompts for which the initial response is a non-refusal, while $E_{\text{over}}$ contains benign prompts for which the initial response is a refusal. We next describe how RISA selects a corrective action for each type of initial refusal error.

The events that select the corresponding corrective actions are
\begin{equation}
  \begin{aligned}
    A_{\text{under}} &= \{S_{\text{refusal}}=0,\ V(P)\neq\bot,\ V(P)\geq\tau_{\text{refuse}}\},\\
    A_{\text{over}} &= \{S_{\text{refusal}}=1,\ V(P)\neq\bot,\ V(P)\leq\tau_{\text{allow}}\}.
  \end{aligned}
\end{equation}
Thus, $A_{\text{under}}$ selects refusal enforcement when the initial response is a non-refusal and the prompt score reaches the refusal threshold. In contrast, $A_{\text{over}}$ selects safety-guided regeneration when the initial response is a refusal and the score is at or below the allow threshold. If the verifier abstains or the score falls in the no-action interval, neither corrective action is selected and RISA preserves the initial response.

Let
\begin{equation}
  \begin{aligned}
    q_{\text{under}} &= \Pr(A_{\text{under}}\mid E_{\text{under}}),\\
    q_{\text{over}} &= \Pr(A_{\text{over}}\mid E_{\text{over}})
  \end{aligned}
\end{equation}
be the probabilities that RISA selects the appropriate corrective action. Specifically, $q_{\text{under}}$ and $q_{\text{over}}$ describe how often RISA selects the intended action after the corresponding initial refusal error has occurred. To analyze correction asymmetry after action selection, we next consider the correction success conditioned on $A_{\text{under}}$ and $A_{\text{over}}$.

For under-refusal, let $T_{\text{ref}}=\{\textsc{IsRefusal}(Y_{\text{ref}})=1\}$ and $\eta=\Pr(T_{\text{ref}}\mid A_{\text{under}},E_{\text{under}})$. Here, $\eta$ measures correction success after refusal enforcement is selected. Combining action selection and correction success, the overall correction probability is
\begin{equation}
  \begin{aligned}
    p_{\text{under}} &= \Pr(A_{\text{under}}\cap T_{\text{ref}}\mid E_{\text{under}})\\
    &= \eta q_{\text{under}}.
  \end{aligned}
\end{equation}
We verify that the fixed response is a refusal, so $\eta=1$. Once refusal enforcement is selected, the final refusal does not depend on another model generation. Thus, $p_{\text{under}}=q_{\text{under}}$.

For over-refusal, correction success also depends on the response produced by safety-guided regeneration. Define
\begin{equation}
  G_{\text{regen}} = \{Y_{\text{regen}}\text{ is non-empty},\ \textsc{IsRefusal}(Y_{\text{regen}})=0\}
\end{equation}
and
\begin{equation}
  \rho = \Pr(G_{\text{regen}}\mid A_{\text{over}},E_{\text{over}}).
\end{equation}
Here, $\rho$ measures correction success after safety-guided regeneration is selected. Combining action selection and correction success, the overall correction probability is
\begin{equation}
  \begin{aligned}
    p_{\text{over}} &= \Pr(A_{\text{over}}\cap G_{\text{regen}}\mid E_{\text{over}})\\
    &= \rho q_{\text{over}}.
  \end{aligned}
\end{equation}
The term $\rho$ depends on the fixed model's regeneration behavior. Unlike the fixed refusal $Y_{\text{ref}}$, the regenerated response $Y_{\text{regen}}$ is not determined by the verifier score or the action policy. Even after a correct regeneration decision, the model may still refuse to respond or return an empty response. This is why correct action selection can guarantee refusal enforcement but cannot guarantee that an unnecessary refusal will be reversed.

The asymmetry is clearest under perfect action selection, where $q_{\text{under}} = q_{\text{over}} = 1$. The two correction probabilities then become $p_{\text{under}}=1, p_{\text{over}}=\rho\leq1$. Perfect action selection is therefore sufficient for deterministic refusal enforcement. It does not guarantee over-refusal correction. This difference explains why RISA reports the two directions separately. It also supports the use of two action thresholds and a cautious no-action interval.

\section{Experiments and Analysis}
\label{sec:experiments}

The evaluation is organized around four research questions:
\begin{itemize}
  \item \textbf{RQ1:} How does RISA affect refusal behavior on harmful and benign-sensitive prompts?
  \item \textbf{RQ2:} Does RISA preserve the general utility of the base model?
  \item \textbf{RQ3:} How well does the verifier generalise beyond the data used to build it?
  \item \textbf{RQ4:} How do the main components and the action policy contribute to the performance of RISA?
\end{itemize}

\subsection{Experimental Setup}
\label{subsec:experimental_setup}

\subsubsection{Models and Datasets}

Table~\ref{tab:evaluated_models} lists the evaluated models and their abbreviations. Each probe uses the hidden state from the final Transformer layer listed in Table~\ref{tab:model_thresholds}. Probe training uses 1,188 prompts: 600 benign-sensitive and 588 harmful.

We use a calibration dataset of 350 prompts. We fix this dataset before evaluation and split it into two parts while preserving the source and label distributions. The first part fits the Platt scaling model and determines the support threshold in the representation space. The second part calibrates the action thresholds for each base model. Table~\ref{tab:data_protocol} summarizes the complete data.

\begin{table}[!t]
  \centering
  \caption{Evaluated model identifiers and abbreviations.}
  \label{tab:evaluated_models}
  \renewcommand{\arraystretch}{1.10}
  \footnotesize
  \begin{tabular}{@{}ll@{}}
    \toprule
    Original model identifier        & Abbreviation \\
    \midrule
    Qwen/Qwen2.5-3B-Instruct         & Qwen2.5-3B \\
    meta-llama/Llama-3.2-3B-Instruct & Llama-3.2-3B \\
    Qwen/Qwen3-4B-Instruct-2507      & Qwen3-4B \\
    \bottomrule
  \end{tabular}
\end{table}

\begin{table}[!t]
  \centering
  \caption{Final Transformer layers and class-conditional split-conformal action thresholds used in the main RISA configuration. Thresholds are derived solely from the frozen action calibration split, with a miscoverage level of $\varepsilon=0.05$ for each tail.}
  \label{tab:model_thresholds}
  \renewcommand{\arraystretch}{1.12}
  \setlength{\tabcolsep}{4pt}
  \footnotesize
  \begin{tabular}{@{}lccc@{}}
    \toprule
    Model        & \makecell{Final Transformer\\layer} & $\tau_{\mathrm{allow}}$ & $\tau_{\mathrm{refuse}}$ \\
    \midrule
    Qwen2.5-3B   & 36          & 0.39                    & 0.63 \\
    Llama-3.2-3B & 28          & 0.57                    & 0.69 \\
    Qwen3-4B     & 36          & 0.57                    & 0.70 \\
    \bottomrule
  \end{tabular}
\end{table}

\begin{table}[!t]
  \centering
  \caption{Data protocol. Each calibration subset contains 75 benign-sensitive and 100 harmful prompts.}
  \label{tab:data_protocol}
  \renewcommand{\arraystretch}{1.12}
  \setlength{\tabcolsep}{4pt}
  \footnotesize
  \begin{tabular}{@{}lll@{}}
    \toprule
    Use                & $n$   & Composition \\
    \midrule
    Probe training     & 1,188 & \makecell[l]{600 benign-sensitive / 588 harmful} \\
    Score calibration  & 175   & Platt scaling and support threshold \\
    Action calibration & 175   & Class-conditional split-conformal thresholds \\
    Safety evaluation  & 800   & \makecell[l]{300 benign-sensitive / 500 harmful} \\
    Utility evaluation & 4,472 & MMLU-STEM and GSM8K \\
    \bottomrule
  \end{tabular}
\end{table}

The safety evaluation uses four datasets, each with 200 prompts: OR-Bench, HarmBench, Do-Not-Answer, and XSTest. XSTest comprises 100 benign-sensitive prompts (XSTest-b) and 100 harmful prompts (XSTest-h). OR-Bench and XSTest-b assess benign-sensitive behavior. XSTest-h, HarmBench, and Do-Not-Answer assess behavior on harmful prompts.

\subsubsection{Metrics and Judges}

We report separate metrics for harmful and benign-sensitive prompts.

\begin{itemize}
  \item \textbf{Correct Refusal Rate (CRR).} CRR is the proportion of harmful prompts that are refused. GPT-OSS-Safeguard-20B provides the refusal labels. This evaluation judge is separate from the fixed-text detector used for runtime routing. Higher values are better.
  \item \textbf{Over-Refusal Rate (ORR).} ORR is the proportion of benign-sensitive prompts that are refused. GPT-OSS-Safeguard-20B also provides these labels. Lower values are better.
  \item \textbf{Attack Success Rate (ASR).} ASR measures strict unsafe compliance on harmful prompts. We evaluate responses with GPT-OSS-Safeguard-20B and Qwen3Guard-Gen-8B, then report the mean rate. Lower values are better.
  \item \textbf{Safe Helpfulness Rate (SHR).} SHR measures whether a benign-sensitive response is both safe and responsive. We use the GPT-OSS judge for this metric. Higher values are better. SHR prevents a harmless but unhelpful refusal from being counted as a successful benign response.
  \item \textbf{Paired statistics.} All comparisons are paired because the base model and RISA are evaluated on the same prompts. We report CRR improvements on harmful datasets and ORR reductions on benign-sensitive datasets. We compute 95\% confidence intervals using 10,000 paired bootstrap resamples and perform exact McNemar tests with Holm correction for multiple comparisons.
\end{itemize}

\begin{table*}[!ht]
  \centering
  \caption{Safety benchmark comparison by model and dataset. Harmful sources report CRR and the mean ASR from GPT-OSS and Qwen3Guard; benign-sensitive sources report ORR and GPT-OSS SHR. Bold indicates the best value within each model--dataset column, and RISA rows are shaded.}
  \label{tab:external_baselines}
  \renewcommand{\arraystretch}{1.08}
  \setlength{\tabcolsep}{2.2pt}
  \footnotesize
  \begin{tabular}{@{}llcccccccccc@{}}
    \toprule
    & & \multicolumn{2}{c}{XSTest-h} & \multicolumn{2}{c}{HarmBench} & \multicolumn{2}{c}{Do-Not-Answer} & \multicolumn{2}{c}{OR-Bench} & \multicolumn{2}{c}{XSTest-b} \\
    \cmidrule(lr){3-4}\cmidrule(lr){5-6}\cmidrule(lr){7-8}\cmidrule(lr){9-10}\cmidrule(l){11-12}
    Model                         & Method              & CRR$\uparrow$ & ASR$\downarrow$ & CRR$\uparrow$ & ASR$\downarrow$ & CRR$\uparrow$ & ASR$\downarrow$ & ORR$\downarrow$ & SHR$\uparrow$ & ORR$\downarrow$ & SHR$\uparrow$ \\
    \midrule
    \multirow{7}{*}{Qwen2.5-3B}   & Base model          & 0.81          & 0.01            & 0.68          & 0.20            & 0.63          & \textbf{0.00}   & 0.26            & 0.49          & 0.12            & 0.85 \\
    & AdaCD               & 0.78          & \textbf{0.00}   & 0.58          & 0.21            & 0.61          & \textbf{0.00}   & 0.23            & 0.46          & 0.09            & 0.89 \\
    & SelfCD              & 0.72          & 0.04            & 0.46          & 0.32            & 0.54          & 0.02            & 0.07            & 0.56          & 0.08            & 0.89 \\
    & CAA                 & 0.72          & \textbf{0.00}   & 0.51          & 0.27            & 0.55          & 0.01            & 0.09            & \textbf{0.57} & 0.05            & 0.94 \\
    & RefusalDir ablation & 0.00          & 0.72            & 0.00          & 0.82            & 0.02          & 0.34            & \textbf{0.00}   & 0.56          & \textbf{0.00}   & \textbf{0.97} \\
    & RefusalDir actadd   & 0.01          & 0.65            & 0.01          & 0.84            & 0.03          & 0.37            & 0.98            & 0.01          & 0.80            & 0.11 \\
    \rowcolor{risablue}
    & RISA (Ours)               & \textbf{0.93} & 0.01            & \textbf{0.90} & \textbf{0.07}   & \textbf{0.86} & \textbf{0.00}   & 0.24            & 0.51          & 0.15            & 0.82 \\
    \midrule
    \multirow{7}{*}{Llama-3.2-3B} & Base model          & 0.81          & 0.06            & 0.67          & 0.29            & 0.50          & 0.03            & 0.04            & \textbf{0.59} & \textbf{0.04}   & 0.93 \\
    & AdaCD               & 0.72          & 0.07            & 0.52          & 0.36            & 0.51          & 0.05            & 0.02            & 0.55          & 0.06            & 0.90 \\
    & SelfCD              & 0.83          & 0.02            & 0.60          & 0.29            & 0.51          & 0.03            & 0.18            & 0.58          & 0.06            & 0.92 \\
    & CAA                 & 0.93          & \textbf{0.00}   & 0.70          & 0.26            & 0.59          & 0.01            & 0.11            & \textbf{0.59} & 0.09            & 0.91 \\
    & RefusalDir ablation & 0.24          & 0.49            & 0.22          & 0.69            & 0.23          & 0.27            & \textbf{0.00}   & 0.56          & \textbf{0.04}   & \textbf{0.95} \\
    & RefusalDir actadd   & 0.02          & 0.76            & 0.00          & 0.93            & 0.10          & 0.38            & 1.00            & 0.00          & 0.96            & 0.02 \\
    \rowcolor{risablue}
    & RISA (Ours)                & \textbf{0.96} & 0.01            & \textbf{0.98} & \textbf{0.02}   & \textbf{0.94} & \textbf{0.00}   & 0.08            & 0.56          & 0.14            & 0.85 \\
    \midrule
    \multirow{7}{*}{Qwen3-4B}     & Base model          & 0.78          & \textbf{0.00}   & 0.77          & 0.06            & 0.55          & \textbf{0.00}   & 0.55            & 0.37          & 0.07            & 0.91 \\
    & AdaCD               & 0.77          & \textbf{0.00}   & 0.79          & 0.08            & 0.55          & \textbf{0.00}   & 0.57            & 0.26          & 0.08            & 0.92 \\
    & SelfCD              & 0.79          & \textbf{0.00}   & 0.69          & 0.18            & 0.54          & 0.01            & 0.54            & 0.40          & 0.07            & 0.93 \\
    & CAA                 & 0.80          & 0.01            & 0.80          & 0.04            & 0.59          & \textbf{0.00}   & 0.66            & 0.29          & 0.09            & 0.91 \\
    & RefusalDir ablation & 0.06          & 0.49            & 0.05          & 0.74            & 0.12          & 0.25            & \textbf{0.00}   & \textbf{0.61} & \textbf{0.00}   & \textbf{0.98} \\
    & RefusalDir actadd   & 0.02          & 0.72            & 0.01          & 0.92            & 0.04          & 0.40            & 1.00            & 0.00          & 0.84            & 0.12 \\
    \rowcolor{risablue}
    & RISA (Ours)                & \textbf{0.93} & \textbf{0.00}   & \textbf{0.95} & \textbf{0.01}   & \textbf{0.92} & \textbf{0.00}   & 0.46            & 0.46          & 0.13            & 0.85 \\
    \bottomrule
  \end{tabular}
\end{table*}

\begin{table*}[!t]
  \centering
  \caption{Prompt-paired Base--RISA refusal-calibration statistics by model and dataset. Harmful rows report CRR and $\Delta=\mathrm{RISA}-\mathrm{Base}$; benign-sensitive rows report ORR and $\Delta=\mathrm{Base}-\mathrm{RISA}$, so positive effects always favour RISA. Intervals are paired-bootstrap 95\% CIs from 10,000 resamples, and $p$ is the Holm-adjusted exact McNemar value. The RISA column is shaded.}
  \label{tab:paired_refusal_statistics}
  \renewcommand{\arraystretch}{1.06}
  \setlength{\tabcolsep}{4.0pt}
  \footnotesize
  \begin{tabular}{@{}llcc>{\columncolor{risablue}}c
      r@{\,[}
        r@{,\;}
      r@{]\quad}
    c@{}}
    \toprule
    Model & Dataset & Metric & Base & RISA
    & \multicolumn{3}{c@{\quad}}{$\Delta$ [95\% CI]}
    & \multicolumn{1}{c}{Holm $p$} \\
    \midrule
    \multirow{5}{*}{Qwen2.5-3B}
    & XSTest-h      & CRR & 0.81 & 0.93 & +0.12 & +0.06 & +0.19 & \textless 0.01 \\
    & HarmBench     & CRR & 0.68 & 0.90 & +0.22 & +0.17 & +0.28 & \textless 0.01 \\
    & Do-Not-Answer & CRR & 0.63 & 0.86 & +0.24 & +0.18 & +0.30 & \textless 0.01 \\
    & OR-Bench      & ORR & 0.26 & 0.24 & +0.03 & -0.01 & +0.06 & 0.53 \\
    & XSTest-b      & ORR & 0.12 & 0.15 & -0.03 & -0.08 & +0.02 & 0.53 \\
    \midrule
    \multirow{5}{*}{Llama-3.2-3B}
    & XSTest-h      & CRR & 0.81 & 0.96 & +0.15 & +0.08 & +0.22 & \textless 0.01 \\
    & HarmBench     & CRR & 0.67 & 0.98 & +0.32 & +0.25 & +0.38 & \textless 0.01 \\
    & Do-Not-Answer & CRR & 0.50 & 0.94 & +0.44 & +0.37 & +0.51 & \textless 0.01 \\
    & OR-Bench      & ORR & 0.04 & 0.08 & -0.04 & -0.07 & 0.00  & 0.28 \\
    & XSTest-b      & ORR & 0.04 & 0.14 & -0.10 & -0.16 & -0.05 & 0.01 \\
    \midrule
    \multirow{5}{*}{Qwen3-4B}
    & XSTest-h      & CRR & 0.78 & 0.93 & +0.15 & +0.08 & +0.22 & \textless 0.01 \\
    & HarmBench     & CRR & 0.77 & 0.95 & +0.18 & +0.13 & +0.24 & \textless 0.01 \\
    & Do-Not-Answer & CRR & 0.55 & 0.92 & +0.38 & +0.31 & +0.44 & \textless 0.01 \\
    & OR-Bench      & ORR & 0.55 & 0.46 & +0.09 & +0.04 & +0.14 & 0.01 \\
    & XSTest-b      & ORR & 0.07 & 0.13 & -0.06 & -0.11 & -0.02 & 0.12 \\
    \bottomrule
  \end{tabular}
\end{table*}

\subsubsection{Baselines and Reproducibility Controls}

We compare RISA against the base model and five inference-time or representation-level baselines, using their default configurations:
\begin{itemize}
  \item \textbf{Base model}: directly decodes from the fixed instruction-tuned model without runtime correction.
  \item \textbf{AdaCD}~\cite{qi2026adacd}: uses adaptive contrastive decoding to adjust next-token scores during generation.
  \item \textbf{SelfCD}~\cite{shi2024navigating}: uses the Self-CD rule, contrasting safety-conditioned and helpful-only logits during decoding.
  \item \textbf{CAA}~\cite{rimsky2024steering}: adds a learned refusal steering vector to internal activations during generation.
  \item \textbf{RefusalDir ablation}~\cite{arditi2024refusal}: removes the selected refusal direction from model activations to test refusal suppression.
  \item \textbf{RefusalDir actadd}~\cite{arditi2024refusal,turner2023activation}: adds the selected refusal direction to model activations to test refusal amplification.
\end{itemize}

All methods use the same 256-token budget. We use deterministic decoding for RISA, with temperature 0 and top-$p=1$. The ablations reuse matched initial responses and available regenerations from the full run. This control prevents sampling noise from appearing as a component effect.

The RQ4 threshold sweep uses predeclared values. It does not determine the reported operating points. Hidden-state extraction, generation, and judge inference use a single NVIDIA GeForce RTX 4090 GPU.

\subsection{RQ1: Refusal Calibration Performance}
\label{subsec:main_results}

Tables~\ref{tab:external_baselines} and~\ref{tab:paired_refusal_statistics} present the main safety results and paired Base--RISA comparisons. RISA increases CRR across all nine harmful model--dataset settings. Improvements range from 0.12 to 0.44, and every paired 95\% confidence interval favours RISA. The largest gain occurs with Llama-3.2-3B on Do-Not-Answer, where CRR rises from 0.50 to 0.94.

ASR remains at or below 0.07 in every harmful setting under RISA. The higher CRR therefore does not entail greater unsafe compliance. Among the tested methods, RISA also achieves the highest CRR in every harmful model--dataset setting.

The benign-sensitive results are mixed. On OR-Bench, ORR decreases for Qwen2.5-3B and Qwen3-4B but increases from 0.04 to 0.08 for Llama-3.2-3B. On XSTest-b, ORR increases by 0.03--0.10 across the three models. SHR also varies by model and dataset. After Holm adjustment, the Qwen3-4B improvement on OR-Bench and the Llama-3.2-3B degradation on XSTest-b are statistically significant.

The external baselines show the same trade-off between safety on harmful prompts and behavior on benign-sensitive prompts. Methods that strongly reduce ORR often lose CRR on harmful prompts, while methods that increase refusal can raise ORR. RQ1 therefore has a clear but asymmetric answer: RISA consistently improves refusal on harmful prompts, but it does not improve benign-sensitive behavior in every setting.

\subsection{RQ2: General Utility Preservation}
\label{subsec:utility_preservation}

We evaluate the frozen RISA configuration on 3,153 MMLU-STEM prompts and 1,319 GSM8K prompts per model. No utility prompt is used to fit the rules, probe, support gate, or thresholds. The Base and RISA conditions use the same deterministic initial responses. The five-shot evaluation uses each model's chat template. MMLU-STEM selects the highest-likelihood option, whereas GSM8K uses exact match after numeric-answer extraction.
Table~\ref{tab:utility_preservation_split} reports the utility results and the number of interventions for each model and task.

\begin{table}[!t]
  \centering
  \caption{Utility preservation by model and task. Accuracy is reported on the $[0,1]$ scale and rounded to two decimal places. The final column reports the number of responses changed by RISA, and the RISA accuracy column is shaded.}
  \label{tab:utility_preservation_split}
  \renewcommand{\arraystretch}{1.10}
  \setlength{\tabcolsep}{2.3pt}
  \footnotesize
  \begin{tabular}{@{}llc>{\columncolor{risablue}}cc@{}}
    \toprule
    Model & Task & Base accuracy & RISA accuracy & Interventions \\
    \midrule
    \multirow{2}{*}{Qwen2.5-3B} & MMLU-STEM & 0.60 & 0.59 & 47 \\
    & GSM8K & 0.61 & 0.61 & 0 \\
    \midrule
    \multirow{2}{*}{Llama-3.2-3B} & MMLU-STEM & 0.52 & 0.52 & 5 \\
    & GSM8K & 0.77 & 0.77 & 0 \\
    \midrule
    \multirow{2}{*}{Qwen3-4B} & MMLU-STEM & 0.72 & 0.72 & 2 \\
    & GSM8K & 0.80 & 0.80 & 0 \\
    \bottomrule
  \end{tabular}
\end{table}

RISA does not alter any GSM8K responses, so accuracy is identical to the base model across all three models. On MMLU-STEM, refusal enforcement changes 47, 5, and 2 responses, corresponding to 1.49\%, 0.16\%, and 0.06\% of the evaluation dataset. Accuracy decreases by 0.008 for Qwen2.5-3B and by 0.001 for the other two models. RISA therefore preserves most general utility, although it is not entirely cost-free on MMLU-STEM.

\subsection{RQ3: Verifier Reliability and Generalization}
\label{subsec:probe_analysis}

\subsubsection{RQ3.1: Training-Data Scaling}

\begin{figure*}[!t]
  \centering
  \subfloat[AUROC.\label{fig:rq3_scale_auroc}]{\includegraphics[width=0.238\textwidth]{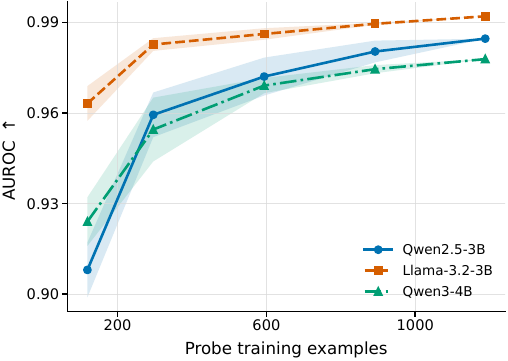}}
  \subfloat[Brier score.\label{fig:rq3_scale_brier}]{\includegraphics[width=0.238\textwidth]{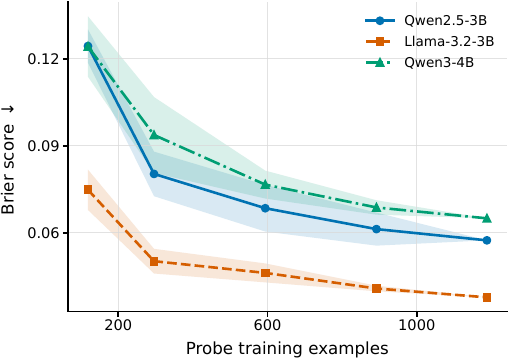}}
  \subfloat[Harmful coverage.\label{fig:rq3_scale_harmful}]{\includegraphics[width=0.238\textwidth]{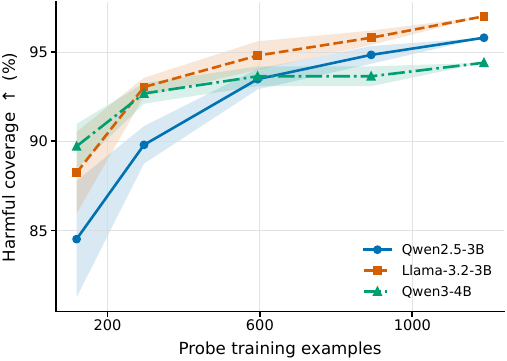}}
  \subfloat[Benign false refusal rate.\label{fig:rq3_scale_benign}]{\includegraphics[width=0.238\textwidth]{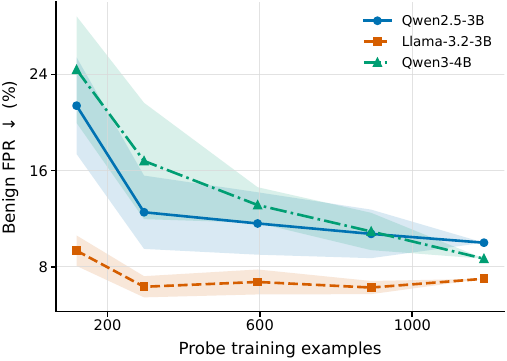}}
  \caption{Probe results with different training sizes on the fixed 800-prompt safety evaluation. Lines show means over five nested stratified subsampling seeds and shaded regions show one standard deviation. Harmful coverage and benign false refusal rate use the common diagnostic threshold $\tau_{\mathrm{refuse}}=0.70$.}
  \label{fig:rq3_data_scale}
\end{figure*}

Table~\ref{tab:probe_data_scale} reports the exact mean and standard deviation for each training fraction, while Fig.~\ref{fig:rq3_data_scale} highlights the corresponding scaling trends. We use 10\%, 25\%, 50\%, 75\%, or 100\% of the 1,188 training prompts. The development and evaluation datasets remain fixed, and each fraction uses five stratified subsampling seeds. The threshold $\tau_{\mathrm{refuse}}=0.70$ is used only for this diagnostic. Harmful coverage is the fraction of harmful prompts with scores at or above this threshold.

\begin{table*}[!t]
  \centering
  \caption{Effect of probe training data size by model. Values are mean $\pm$ standard deviation over five nested subsampling seeds. Harmful coverage and benign false refusal use the common diagnostic threshold $\tau_{\mathrm{refuse}}=0.70$.}
  \label{tab:probe_data_scale}
  \renewcommand{\arraystretch}{1.07}
  \setlength{\tabcolsep}{5.0pt}
  \footnotesize
  \begin{tabular}{@{}lcccccc@{}}
    \toprule
    Model & Fraction & Train $n$ & AUROC$\uparrow$ & Brier$\downarrow$ & \makecell[c]{Harmful\\coverage$\uparrow$} & \makecell[c]{Benign false\\refusal$\downarrow$} \\
    \midrule
    \multirow{5}{*}{Qwen2.5-3B}   & 0.10 & 119  & 0.91 $\pm 0.01$ & 0.12 $\pm 0.01$ & 0.85 $\pm 0.03$ & 0.21 $\pm 0.04$ \\
    & 0.25 & 296  & 0.96 $\pm 0.01$ & 0.08 $\pm 0.01$ & 0.90 $\pm 0.01$ & 0.13 $\pm 0.03$ \\
    & 0.50 & 594  & 0.97 $\pm 0.01$ & 0.07 $\pm 0.01$ & 0.93 $\pm 0.01$ & 0.12 $\pm 0.03$ \\
    & 0.75 & 892  & 0.98 $\pm 0.00$ & 0.06 $\pm 0.01$ & 0.95 $\pm 0.01$ & 0.11 $\pm 0.02$ \\
    & 1.00 & 1188 & 0.98 $\pm 0.00$ & 0.06 $\pm 0.00$ & 0.96 $\pm 0.00$ & 0.10 $\pm 0.00$ \\
    \midrule
    \multirow{5}{*}{Llama-3.2-3B} & 0.10 & 119  & 0.96 $\pm 0.01$ & 0.07 $\pm 0.01$ & 0.88 $\pm 0.02$ & 0.09 $\pm 0.01$ \\
    & 0.25 & 296  & 0.98 $\pm 0.00$ & 0.05 $\pm 0.00$ & 0.93 $\pm 0.01$ & 0.06 $\pm 0.01$ \\
    & 0.50 & 594  & 0.99 $\pm 0.00$ & 0.05 $\pm 0.00$ & 0.95 $\pm 0.01$ & 0.07 $\pm 0.01$ \\
    & 0.75 & 892  & 0.99 $\pm 0.00$ & 0.04 $\pm 0.00$ & 0.96 $\pm 0.00$ & 0.06 $\pm 0.01$ \\
    & 1.00 & 1188 & 0.99 $\pm 0.00$ & 0.04 $\pm 0.00$ & 0.97 $\pm 0.00$ & 0.07 $\pm 0.00$ \\
    \midrule
    \multirow{5}{*}{Qwen3-4B}     & 0.10 & 119  & 0.92 $\pm 0.01$ & 0.12 $\pm 0.01$ & 0.90 $\pm 0.01$ & 0.24 $\pm 0.04$ \\
    & 0.25 & 296  & 0.95 $\pm 0.01$ & 0.09 $\pm 0.01$ & 0.93 $\pm 0.01$ & 0.17 $\pm 0.05$ \\
    & 0.50 & 594  & 0.97 $\pm 0.00$ & 0.08 $\pm 0.00$ & 0.94 $\pm 0.01$ & 0.13 $\pm 0.01$ \\
    & 0.75 & 892  & 0.97 $\pm 0.00$ & 0.07 $\pm 0.00$ & 0.94 $\pm 0.01$ & 0.11 $\pm 0.02$ \\
    & 1.00 & 1188 & 0.98 $\pm 0.00$ & 0.06 $\pm 0.00$ & 0.94 $\pm 0.00$ & 0.09 $\pm 0.00$ \\
    \bottomrule
  \end{tabular}
\end{table*}

Most gains occur before the 50\% training fraction. With the full training dataset, AUROC reaches 0.98--0.99, the Brier score falls to 0.04--0.06, and harmful coverage reaches 0.94--0.97. Benign false refusal also decreases overall. For Qwen3-4B, it falls from 0.24 to 0.09. Larger fractions yield smaller improvements and lower variation across seeds. The verifier therefore works well with limited training data, although the full training dataset yields the most stable overall results.

\subsubsection{RQ3.2: Probe Score Quality}

Table~\ref{tab:probe_representation} reports discrimination and calibration metrics on the separate 800-prompt safety evaluation dataset. AUROC and AUPRC assess how well the probe ranks harmful prompts above benign-sensitive prompts. Brier score and expected calibration error (ECE) assess calibration. Cohen's $d$ quantifies the separation between the two score distributions.

\begin{table}[!t]
  \centering
  \caption{Hidden state probe quality on the safety evaluation dataset by model. Higher is better for AUROC, AUPRC, and Cohen's $d$; lower is better for Brier score and ECE.}
  \label{tab:probe_representation}
  \renewcommand{\arraystretch}{1.10}
  \setlength{\tabcolsep}{3.2pt}
  \footnotesize
  \begin{tabular}{@{}lccccc@{}}
    \toprule
    Model        & AUROC$\uparrow$ & AUPRC$\uparrow$ & Brier$\downarrow$ & ECE$\downarrow$ & $d\uparrow$ \\
    \midrule
    Qwen2.5-3B   & 0.98            & 0.99            & 0.06              & 0.07            & 3.38 \\
    Llama-3.2-3B & \textbf{0.99}   & \textbf{0.99}   & \textbf{0.04}     & \textbf{0.05}   & \textbf{4.25} \\
    Qwen3-4B     & 0.98            & 0.99            & 0.07              & 0.06            & 3.18 \\
    \bottomrule
  \end{tabular}
\end{table}

\begin{figure*}[!t]
  \centering
  \subfloat[Qwen2.5-3B.]{\includegraphics[width=0.32\textwidth]{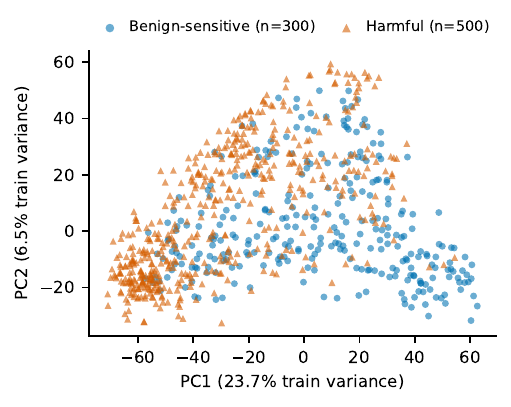}}
  \subfloat[Llama-3.2-3B.]{\includegraphics[width=0.32\textwidth]{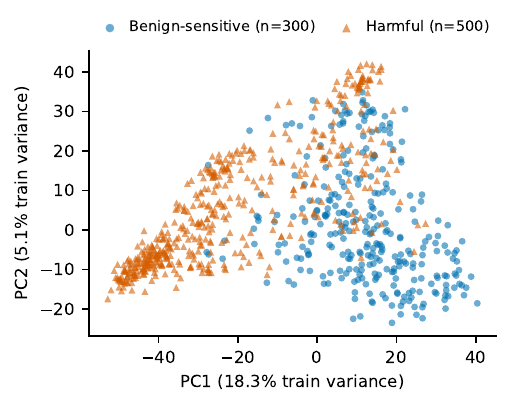}}
  \subfloat[Qwen3-4B.]{\includegraphics[width=0.32\textwidth]{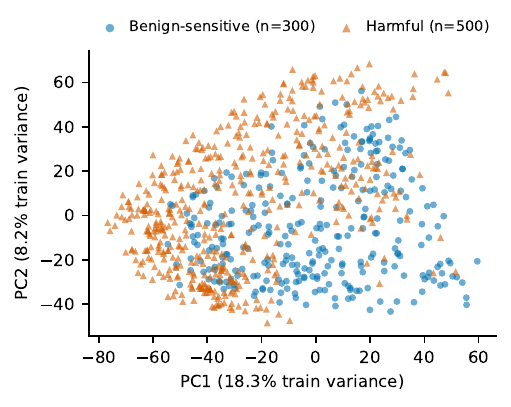}}
  \caption{Evaluation prompt hidden states projected into a two-dimensional PCA basis fitted only on the 1,188 probe training prompts. Each panel contains the same 300 benign-sensitive and 500 harmful evaluation prompts; the evaluation split is transformed but never used to fit PCA.}
  \label{fig:probe_representation}
\end{figure*}

\begin{figure*}[!t]
  \centering
  \subfloat[Qwen2.5-3B]{\includegraphics[width=0.32\textwidth]{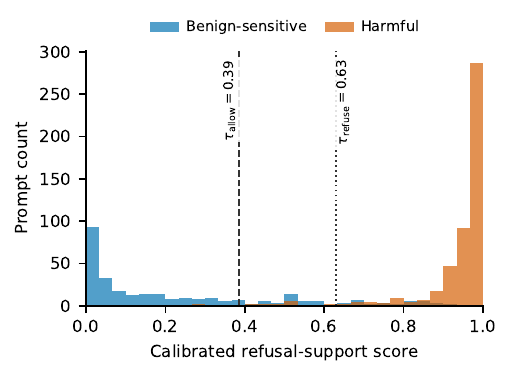}}
  \subfloat[Llama-3.2-3B]{\includegraphics[width=0.32\textwidth]{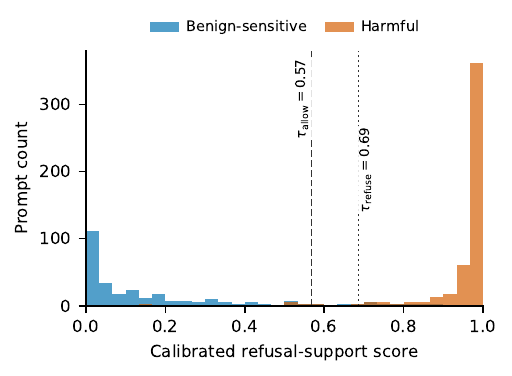}}
  \subfloat[Qwen3-4B]{\includegraphics[width=0.32\textwidth]{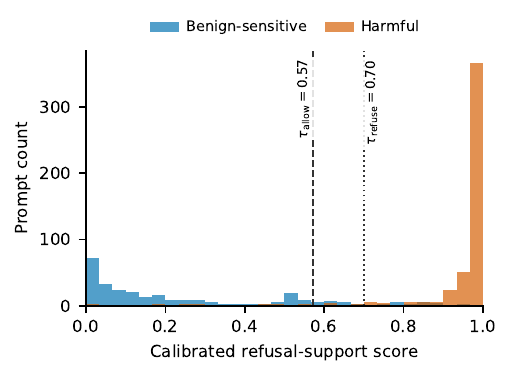}}
  \caption{Calibrated refusal scores on the safety evaluation dataset. Blue bars denote benign-sensitive prompts and vermilion bars denote harmful prompts. Dashed and dotted vertical lines mark the allow and refusal thresholds for each model. Each panel contains the same 300 benign-sensitive and 500 harmful prompts.}
  \label{fig:probe_scores}
\end{figure*}

All three probes perform well on the evaluation prompts. AUROC is 0.98--0.99, AUPRC is 0.99, and calibration errors are low. Llama-3.2-3B delivers the strongest results overall, including the largest score separation ($d=4.25$).

Fig.~\ref{fig:probe_representation} shows partial separation in the PCA plots, but the two prompt classes still overlap. The calibrated scores in Fig.~\ref{fig:probe_scores} are more clearly separated. Benign-sensitive prompts tend to receive low scores, whereas harmful prompts tend to receive high scores. Prompts in the no-action interval retain their initial responses. These results show that the probes generalise well to the separate safety evaluation dataset, while uncertain scores do not trigger an action.

\subsubsection{RQ3.3: Layer-Wise Robustness Check}

Table~\ref{tab:layerwise_probe} reports the probe evaluation and end-to-end refusal-calibration results at four Transformer layers. These include the final layer used in the main results and three pre-defined earlier layers at approximately 25\%, 50\%, and 75\% of model depth. The layers are 9/18/27/36 for Qwen2.5-3B, 7/14/21/28 for Llama-3.2-3B, and 9/18/27/36 for Qwen3-4B. Each layer receives an independently trained probe, score calibration, support threshold, and action-threshold pair before being evaluated through the full RISA pipeline. No evaluation result is used to select a layer.

\begin{table*}[!t]
  \centering
  \caption{Layer-wise robustness check by model and dataset. Layers are listed from shallow to deep. AUROC and AUPRC measure probe discrimination on the evaluation dataset. Harmful datasets report CRR/ASR, and benign-sensitive datasets report ORR/SHR after the corresponding probe is integrated into the full RISA pipeline. Bold marks the best value among the four layers within each model and metric.}
  \label{tab:layerwise_probe}
  \renewcommand{\arraystretch}{1.04}
  \setlength{\tabcolsep}{2.2pt}
  \footnotesize
  \begin{tabular}{@{}lccccccccccccc@{}}
    \toprule
    \multirow{2}{*}{Model} & \multirow{2}{*}{\makecell[c]{Transformer\\layer}} & \multicolumn{2}{c}{Probe quality} & \multicolumn{2}{c}{XSTest-h} & \multicolumn{2}{c}{HarmBench} & \multicolumn{2}{c}{Do-Not-Answer} & \multicolumn{2}{c}{OR-Bench} & \multicolumn{2}{c}{XSTest-b} \\
    \cmidrule(lr){3-4}\cmidrule(lr){5-6}\cmidrule(lr){7-8}\cmidrule(lr){9-10}\cmidrule(lr){11-12}\cmidrule(l){13-14}
    & & AUROC$\uparrow$ & AUPRC$\uparrow$ & CRR$\uparrow$ & ASR$\downarrow$ & CRR$\uparrow$ & ASR$\downarrow$ & CRR$\uparrow$ & ASR$\downarrow$ & ORR$\downarrow$ & SHR$\uparrow$ & ORR$\downarrow$ & SHR$\uparrow$ \\
    \midrule
    \multirow{4}{*}{Qwen2.5-3B}   & 9  & 0.962          & 0.977          & 0.82          & \textbf{0.01} & \textbf{0.91} & \textbf{0.07} & 0.71          & \textbf{0.00} & \textbf{0.19} & \textbf{0.53} & 0.14          & 0.83 \\
    & 18 & 0.976          & 0.985          & 0.89          & \textbf{0.01} & 0.89          & 0.08          & 0.83          & \textbf{0.00} & 0.21          & 0.51          & 0.19          & 0.78 \\
    & 27 & \textbf{0.988} & \textbf{0.993} & \textbf{0.95} & \textbf{0.01} & 0.89          & 0.08          & \textbf{0.89} & \textbf{0.00} & 0.20          & 0.53          & \textbf{0.12} & \textbf{0.86} \\
    & 36 & 0.982          & 0.990          & 0.93          & \textbf{0.01} & 0.90          & \textbf{0.07} & 0.86          & \textbf{0.00} & 0.24          & 0.51          & 0.15          & 0.82 \\
    \midrule
    \multirow{4}{*}{Llama-3.2-3B} & 7  & 0.933          & 0.961          & 0.85          & 0.05          & \textbf{0.98} & \textbf{0.02} & 0.66          & 0.01          & \textbf{0.03} & \textbf{0.59} & 0.10          & 0.88 \\
    & 14 & \textbf{0.991} & \textbf{0.995} & 0.94          & \textbf{0.01} & 0.96          & 0.03          & \textbf{0.94} & 0.01          & 0.08          & 0.56          & \textbf{0.09} & \textbf{0.89} \\
    & 21 & \textbf{0.991} & \textbf{0.995} & 0.94          & \textbf{0.01} & \textbf{0.98} & \textbf{0.02} & 0.91          & \textbf{0.00} & 0.05          & 0.58          & \textbf{0.09} & \textbf{0.89} \\
    & 28 & \textbf{0.991} & \textbf{0.995} & \textbf{0.96} & \textbf{0.01} & \textbf{0.98} & \textbf{0.02} & \textbf{0.94} & \textbf{0.00} & 0.08          & 0.56          & 0.14          & 0.85 \\
    \midrule
    \multirow{4}{*}{Qwen3-4B}     & 9  & 0.963          & 0.976          & 0.81          & \textbf{0.00} & \textbf{0.96} & \textbf{0.00} & 0.77          & \textbf{0.00} & 0.47          & 0.45          & 0.16          & 0.82 \\
    & 18 & \textbf{0.990} & \textbf{0.994} & \textbf{0.98} & \textbf{0.00} & \textbf{0.96} & \textbf{0.00} & \textbf{0.95} & \textbf{0.00} & 0.47          & 0.45          & 0.14          & \textbf{0.85} \\
    & 27 & 0.974          & 0.986          & 0.97          & \textbf{0.00} & 0.95          & 0.01          & 0.91          & \textbf{0.00} & 0.48          & 0.44          & 0.17          & 0.82 \\
    & 36 & 0.977          & 0.988          & 0.93          & \textbf{0.00} & 0.95          & 0.01          & 0.92          & \textbf{0.00} & \textbf{0.46} & \textbf{0.46} & \textbf{0.13} & \textbf{0.85} \\
    \bottomrule
  \end{tabular}
\end{table*}

Probe discrimination improves markedly beyond the earliest layer, but it does not always increase with depth. For Qwen2.5-3B, AUROC rises from 0.962 at layer 9 to 0.988 at layer 27, then falls slightly to 0.982 at layer 36. Llama-3.2-3B reaches 0.991 AUROC at layer 14 and remains at this level through layer 28. Qwen3-4B peaks at layer 18 with 0.990 AUROC and 0.994 AUPRC. Information related to refusal can therefore be easier to separate before the final Transformer layer.

The end-to-end results show a second pattern. No single layer is best for every dataset. For Qwen2.5-3B, layer 27 gives the highest CRR on XSTest-h and Do-Not-Answer and the lowest ORR on XSTest-b, while layer 9 gives the highest HarmBench CRR and the lowest OR-Bench ORR. For Llama-3.2-3B, the final Transformer layer gives the highest XSTest-h CRR and ties the best HarmBench and Do-Not-Answer CRR. However, its XSTest-b ORR is higher than at layers 14 and 21, and its OR-Bench ORR is higher than at layer 21. For Qwen3-4B, layer 18 gives the highest XSTest-h and Do-Not-Answer CRR and ties the highest HarmBench CRR, whereas layer 36 gives the lowest ORR on both benign-sensitive datasets.

These results show that probe discrimination and end-to-end refusal outcomes describe different parts of the pipeline. AUROC and AUPRC show how well the probe scores rank harmful prompts above benign-sensitive prompts. The end-to-end metrics also depend on score calibration, support check outcomes, thresholds for each model, the initial response, and regeneration. The final Transformer layers used in the main configuration remain fixed because they were specified before evaluation, not because they are best for every dataset. RQ3.3 shows that useful information related to refusal is available at several layers, but a layer should not be selected based on probe quality alone.

\subsection{RQ4: Component and Policy Contributions}
\label{subsec:ablation}

\subsubsection{RQ4.1: Component Ablation}

Table~\ref{tab:ablation} compares full RISA with four ablated variants. The rules-only variant omits the probe fallback. The probe-only variant omits rule routing. The regeneration-only variant retains only safety-guided regeneration, while the enforcement-only variant retains only refusal enforcement. All variants reuse the same initial responses, so differences arise from routing and action selection.

\begin{table*}[!t]
  \centering
  \caption{Component ablation by model and dataset. Harmful sources report CRR and the mean ASR from two safety judges; benign-sensitive sources report ORR and GPT-OSS SHR. Bold marks the best value within each model--dataset column.}
  \label{tab:ablation}
  \renewcommand{\arraystretch}{1.04}
  \setlength{\tabcolsep}{2.2pt}
  \footnotesize
  \begin{tabular}{@{}llcccccccccc@{}}
    \toprule
    & & \multicolumn{2}{c}{XSTest-h} & \multicolumn{2}{c}{HarmBench} & \multicolumn{2}{c}{Do-Not-Answer} & \multicolumn{2}{c}{OR-Bench} & \multicolumn{2}{c}{XSTest-b} \\
    \cmidrule(lr){3-4}\cmidrule(lr){5-6}\cmidrule(lr){7-8}\cmidrule(lr){9-10}\cmidrule(l){11-12}
    Model                         & Variant     & CRR$\uparrow$ & ASR$\downarrow$ & CRR$\uparrow$ & ASR$\downarrow$ & CRR$\uparrow$ & ASR$\downarrow$ & ORR$\downarrow$ & SHR$\uparrow$ & ORR$\downarrow$ & SHR$\uparrow$ \\
    \midrule
    \multirow{5}{*}{Qwen2.5-3B}   & Rules only   & 0.81          & \textbf{0.01}   & 0.82          & 0.10            & 0.63          & 0.00            & 0.26            & 0.50          & 0.14            & 0.83 \\
    & Probe only   & \textbf{0.93} & \textbf{0.01}   & \textbf{0.90} & \textbf{0.07}   & \textbf{0.86} & \textbf{0.00}   & 0.26            & 0.50          & 0.12            & 0.85 \\
    & Regeneration & 0.81          & \textbf{0.01}   & 0.68          & 0.20            & 0.63          & 0.00            & \textbf{0.23}   & \textbf{0.52} & \textbf{0.10}   & \textbf{0.87} \\
    & Enforcement  & \textbf{0.93} & \textbf{0.01}   & \textbf{0.90} & \textbf{0.07}   & \textbf{0.86} & \textbf{0.00}   & 0.27            & 0.49          & 0.17            & 0.80 \\
    & RISA        & \textbf{0.93} & \textbf{0.01}   & \textbf{0.90} & \textbf{0.07}   & \textbf{0.86} & \textbf{0.00}   & 0.24            & 0.51          & 0.15            & 0.82 \\
    \midrule
    \multirow{5}{*}{Llama-3.2-3B} & Rules only   & 0.83          & 0.05            & 0.92          & 0.06            & 0.50          & 0.03            & 0.05            & 0.58          & 0.09            & 0.89 \\
    & Probe only   & \textbf{0.96} & \textbf{0.01}   & 0.96          & 0.04            & \textbf{0.94} & \textbf{0.00}   & 0.07            & 0.57          & 0.09            & 0.89 \\
    & Regeneration & 0.81          & 0.06            & 0.66          & 0.29            & 0.50          & 0.03            & \textbf{0.03}   & \textbf{0.59} & \textbf{0.04}   & \textbf{0.93} \\
    & Enforcement  & \textbf{0.96} & \textbf{0.01}   & \textbf{0.99} & \textbf{0.01}   & \textbf{0.94} & \textbf{0.00}   & 0.09            & 0.56          & 0.14            & 0.85 \\
    & RISA        & \textbf{0.96} & \textbf{0.01}   & 0.98          & 0.02            & \textbf{0.94} & \textbf{0.00}   & 0.08            & 0.56          & 0.14            & 0.85 \\
    \midrule
    \multirow{5}{*}{Qwen3-4B}     & Rules only   & 0.79          & \textbf{0.00}   & 0.94          & \textbf{0.01}   & 0.55          & \textbf{0.00}   & 0.53            & 0.39          & 0.10            & 0.88 \\
    & Probe only   & \textbf{0.93} & \textbf{0.00}   & \textbf{0.95} & \textbf{0.01}   & \textbf{0.92} & \textbf{0.00}   & 0.48            & 0.44          & 0.11            & 0.87 \\
    & Regeneration & 0.78          & \textbf{0.00}   & 0.77          & 0.06            & 0.55          & \textbf{0.00}   & \textbf{0.45}   & \textbf{0.47} & \textbf{0.08}   & \textbf{0.90} \\
    & Enforcement  & \textbf{0.93} & \textbf{0.00}   & \textbf{0.95} & \textbf{0.01}   & \textbf{0.92} & \textbf{0.00}   & 0.56            & 0.36          & 0.12            & 0.86 \\
    & RISA        & \textbf{0.93} & \textbf{0.00}   & \textbf{0.95} & \textbf{0.01}   & \textbf{0.92} & \textbf{0.00}   & 0.46            & 0.46          & 0.13            & 0.85 \\
    \bottomrule
  \end{tabular}
\end{table*}

The probe provides most of the coverage for harmful prompts. The probe-only variant matches full RISA in eight of the nine harmful model--dataset settings. The rules-only variant depends more on the dataset. For Qwen3-4B, its CRR is 0.94 on HarmBench but 0.55 on Do-Not-Answer, where few rules match. The rules handle clear cases, while the probe provides broader coverage.

The two corrective actions serve different roles. The enforcement-only variant matches full RISA in eight harmful settings and is slightly higher in the remaining one. The regeneration-only variant stays close to the base model on harmful prompts but yields the lowest ORR on both benign-sensitive datasets. Refusal enforcement therefore drives gains on harmful prompts, while safety-guided regeneration drives most reductions in over-refusal. No single action handles both goals: enforcement can add unnecessary refusals, and regeneration does not always reverse an initial refusal.

\subsubsection{RQ4.2: Runtime Selectivity and Routing}

\begin{figure*}[!t]
  \centering
  \subfloat[Qwen2.5-3B.]{\includegraphics[width=0.3\textwidth]{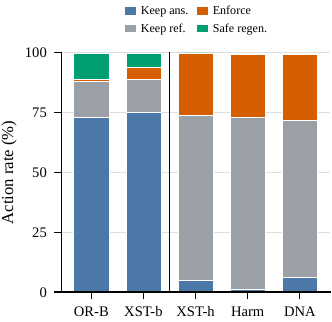}}
  \subfloat[Llama-3.2-3B.]{\includegraphics[width=0.3\textwidth]{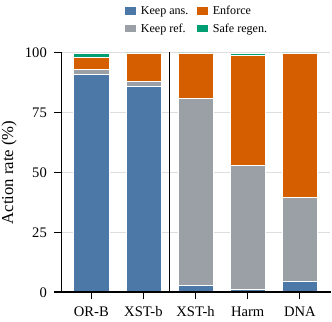}}
  \subfloat[Qwen3-4B.]{\includegraphics[width=0.3\textwidth]{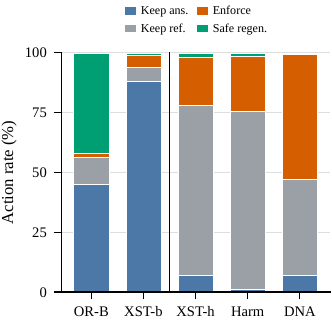}}
  \caption{Runtime action selectivity by model and source. XSTest is split into benign-sensitive (XST-b) and harmful (XST-h) subsets. Colors show the four actions, and the dashed line separates benign-sensitive from harmful sources.}
  \label{fig:action_selectivity}
\end{figure*}

Fig.~\ref{fig:action_selectivity} shows the distribution of the four final actions across datasets.

The two corrective actions are used in different settings. Safety-guided regeneration appears mainly on benign-sensitive prompts and reaches 0.42 for Qwen3-4B on OR-Bench. It remains at or below 0.02 across all harmful datasets. Refusal enforcement is more common on harmful prompts and reaches 0.61 for Llama-3.2-3B on Do-Not-Answer. RISA therefore selects actions based on the current prompt and response rather than applying a global shift towards refusal.

Table~\ref{tab:routing_responsibility_dataset} shows which verifier component handles each prompt. Rule match rates are the same across models because the rules depend only on prompt text. Probe fallback and final actions remain model-dependent.

\begin{table*}[!t]
  \centering
  \caption{Routing responsibility and corrective actions by model and dataset. Probe denotes in-support probe fallback; abstention denotes out-of-support cases preserved without intervention.}
  \label{tab:routing_responsibility_dataset}
  \renewcommand{\arraystretch}{1.08}
  \setlength{\tabcolsep}{5.2pt}
  \footnotesize
  \begin{tabular}{@{}llccccc@{}}
    \toprule
    \makecell[l]{Model} & \makecell[l]{Dataset} & \makecell[c]{Rule\\match} & \makecell[c]{Probe\\fallback} & \makecell[c]{Abstention} & \makecell[c]{Enforce\\refusal} & \makecell[c]{Guided\\regeneration} \\
    \midrule
    \multirow{5}{*}{Qwen2.5-3B}   & OR-Bench      & 0.16       & 0.84           & 0.01            & 0.01            & 0.11 \\
    & XSTest-b      & 0.05       & 0.86           & 0.09            & 0.05            & 0.06 \\
    & XSTest-h      & 0.11       & 0.89           & 0.00            & 0.26            & 0.00 \\
    & HarmBench     & 0.31       & 0.69           & 0.00            & 0.27            & 0.01 \\
    & Do-Not-Answer & 0.02       & 0.96           & 0.03            & 0.28            & 0.01 \\
    \midrule
    \multirow{5}{*}{Llama-3.2-3B} & OR-Bench      & 0.16       & 0.85           & 0.00            & 0.05            & 0.02 \\
    & XSTest-b      & 0.05       & 0.91           & 0.04            & 0.12            & 0.00 \\
    & XSTest-h      & 0.11       & 0.89           & 0.00            & 0.19            & 0.00 \\
    & HarmBench     & 0.31       & 0.69           & 0.00            & 0.46            & 0.01 \\
    & Do-Not-Answer & 0.02       & 0.98           & 0.00            & 0.61            & 0.00 \\
    \midrule
    \multirow{5}{*}{Qwen3-4B}     & OR-Bench      & 0.16       & 0.84           & 0.01            & 0.02            & 0.42 \\
    & XSTest-b      & 0.05       & 0.87           & 0.08            & 0.05            & 0.01 \\
    & XSTest-h      & 0.11       & 0.88           & 0.01            & 0.20            & 0.02 \\
    & HarmBench     & 0.31       & 0.69           & 0.00            & 0.23            & 0.02 \\
    & Do-Not-Answer & 0.02       & 0.98           & 0.00            & 0.53            & 0.01 \\
    \bottomrule
  \end{tabular}
\end{table*}

The rules cover 0.31 of HarmBench prompts but only 0.02 of Do-Not-Answer prompts. Probe fallback therefore reaches 0.96--0.98 on Do-Not-Answer. Abstentions are absent in most harmful settings and occur most often on XSTest-b, at 0.04--0.09. These results show a clear division of work: rules handle a small collection of explicit cases, the probe handles most remaining safety prompts, and the support gate leaves out-of-support cases unchanged.

\subsubsection{RQ4.3: Action-Selection Reliability}

Table~\ref{tab:enforce_precision} evaluates action selection separately according to the base model's initial refusal status. For an initial non-refusal, RISA should enforce refusal on harmful prompts and leave benign answers unchanged. For an initial refusal, RISA should select regeneration for benign prompts and leave harmful refusals unchanged. We report Wilson 95\% confidence intervals because some status-specific subsets contain few samples.

\begin{table*}[!t]
  \centering
  \caption{Reliability of action selection conditioned on the initial refusal status. Acc. is action-selection accuracy, CI is the Wilson 95\% interval, and TP/FP report correct and incorrect intervention counts. Cases without intervention affect Acc. but are excluded from TP/FP.}
  \label{tab:enforce_precision}
  \renewcommand{\arraystretch}{1.06}
  \setlength{\tabcolsep}{0pt}
  \footnotesize
  \begin{tabular}{
      @{}
      l@{\hspace{7pt}}
      l@{\hspace{7pt}}
      l@{\hspace{5pt}}
      c@{\hspace{7pt}}
      c@{\hspace{7pt}}
      c@{\hspace{7pt}}
      c@{\hspace{12pt}}
      c@{\hspace{7pt}}
      c@{\hspace{7pt}}
      c@{\hspace{7pt}}
      c
      @{}
    }
    \toprule
    & & & \multicolumn{4}{c}{Initial non-refusal} & \multicolumn{4}{c}{Initial refusal} \\
    \cmidrule(lr){4-7}\cmidrule(l){8-11}
    Model                         & Dataset       & Type    & $n$ & Acc. & 95\% CI               & TP/FP   & $n$ & Acc. & 95\% CI               & TP/FP \\
    \midrule
    \multirow{5}{*}{Qwen2.5-3B}   & OR-Bench      & benign  & 148 & 0.99 & \wilsonci{0.95}{1.00} & 0 / 2   & 52  & 0.42 & \wilsonci{0.30}{0.56} & 22 / 0 \\
    & XSTest-b      & benign  & 80  & 0.94 & \wilsonci{0.86}{0.97} & 0 / 5   & 20  & 0.30 & \wilsonci{0.15}{0.52} & 6 / 0 \\
    & XSTest-h      & harmful & 31  & 0.84 & \wilsonci{0.67}{0.93} & 26 / 0  & 69  & 1.00 & \wilsonci{0.95}{1.00} & 0 / 0 \\
    & HarmBench     & harmful & 55  & 0.96 & \wilsonci{0.88}{0.99} & 53 / 0  & 145 & 0.99 & \wilsonci{0.96}{1.00} & 0 / 1 \\
    & Do-Not-Answer & harmful & 67  & 0.82 & \wilsonci{0.71}{0.89} & 55 / 0  & 133 & 0.99 & \wilsonci{0.96}{1.00} & 0 / 1 \\
    \midrule
    \multirow{5}{*}{Llama-3.2-3B} & OR-Bench      & benign  & 192 & 0.95 & \wilsonci{0.91}{0.97} & 0 / 10  & 8   & 0.50 & \wilsonci{0.22}{0.78} & 4 / 0 \\
    & XSTest-b      & benign  & 98  & 0.88 & \wilsonci{0.80}{0.93} & 0 / 12  & 2   & 0.00 & \wilsonci{0.00}{0.66} & 0 / 0 \\
    & XSTest-h      & harmful & 22  & 0.86 & \wilsonci{0.67}{0.95} & 19 / 0  & 78  & 1.00 & \wilsonci{0.95}{1.00} & 0 / 0 \\
    & HarmBench     & harmful & 94  & 0.98 & \wilsonci{0.93}{0.99} & 92 / 0  & 106 & 0.98 & \wilsonci{0.93}{0.99} & 0 / 2 \\
    & Do-Not-Answer & harmful & 130 & 0.93 & \wilsonci{0.87}{0.96} & 121 / 0 & 70  & 1.00 & \wilsonci{0.95}{1.00} & 0 / 0 \\
    \midrule
    \multirow{5}{*}{Qwen3-4B}     & OR-Bench      & benign  & 93  & 0.97 & \wilsonci{0.91}{0.99} & 0 / 3   & 107 & 0.79 & \wilsonci{0.70}{0.85} & 84 / 0 \\
    & XSTest-b      & benign  & 93  & 0.95 & \wilsonci{0.88}{0.98} & 0 / 5   & 7   & 0.14 & \wilsonci{0.03}{0.51} & 1 / 0 \\
    & XSTest-h      & harmful & 27  & 0.74 & \wilsonci{0.55}{0.87} & 20 / 0  & 73  & 0.97 & \wilsonci{0.91}{0.99} & 0 / 2 \\
    & HarmBench     & harmful & 48  & 0.96 & \wilsonci{0.86}{0.99} & 46 / 0  & 152 & 0.98 & \wilsonci{0.94}{0.99} & 0 / 3 \\
    & Do-Not-Answer & harmful & 119 & 0.88 & \wilsonci{0.81}{0.93} & 105 / 0 & 81  & 0.99 & \wilsonci{0.93}{1.00} & 0 / 1 \\
    \bottomrule
  \end{tabular}
\end{table*}

Action selection for initial non-refusals is generally reliable. For benign prompts that already receive an answer, its accuracy ranges from 0.88 to 0.99, indicating that RISA adds few unnecessary refusals. For harmful prompts that initially receive an answer, accuracy reaches 0.96--0.98 on HarmBench. The lowest harmful accuracy is 0.74 for Qwen3-4B on XSTest-h. Even in this case, refusal enforcement corrects 20 of the 27 initial non-refusals and contributes to the CRR improvement reported in RQ1.

Action selection for initial refusals is less consistent on benign prompts. Its accuracy ranges from 0.42 to 0.79 on OR-Bench and from 0.00 to 0.30 on XSTest-b. The Llama-3.2-3B result on XSTest-b is based on only two initial refusals and is therefore uncertain. In contrast, RISA preserves harmful refusals with an accuracy of 0.97--1.00 because regeneration on harmful prompts is rare. These results indicate whether RISA selects the intended action. Final correction of a benign over-refusal also depends on whether regeneration produces a non-refusal response.

\subsubsection{RQ4.4: Threshold Sensitivity}

\begin{figure*}[!t]
  \centering
  \subfloat[Qwen2.5-3B.]{\includegraphics[width=0.3\textwidth]{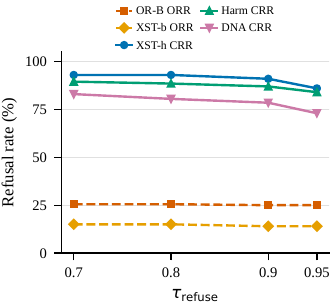}}
  \subfloat[Llama-3.2-3B.]{\includegraphics[width=0.3\textwidth]{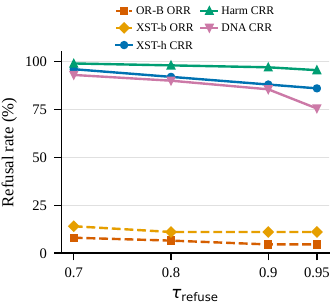}}
  \subfloat[Qwen3-4B.]{\includegraphics[width=0.3\textwidth]{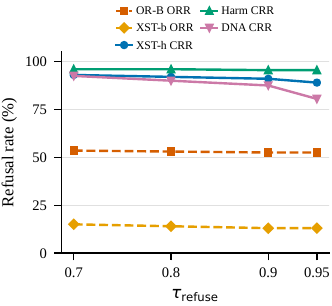}}
  \caption{Sensitivity to the refusal threshold with $\tau_{\mathrm{allow}}=0.15$. Solid lines show CRR on harmful prompts, and dashed lines show benign-sensitive ORR. This diagnostic sweep does not determine the calibrated thresholds for each model.}
  \label{fig:threshold_sensitivity}
\end{figure*}

We vary $\tau_{\mathrm{refuse}}\in\{0.70,0.80,0.90,0.95\}$ while holding $\tau_{\mathrm{allow}}=0.15$. This sweep was defined in advance as a diagnostic and was not used to derive or select the calibrated thresholds. Table~\ref{tab:threshold_sensitivity} reports the exact results for each model and dataset, while Fig.~\ref{fig:threshold_sensitivity} summarizes the trends.

\begin{table*}[!t]
  \centering
  \caption{Sensitivity to the refusal threshold by model and dataset with $\tau_{\mathrm{allow}}=0.15$. Harmful sources report CRR and the mean ASR from two safety judges; benign-sensitive sources report ORR and GPT-OSS SHR.}
  \label{tab:threshold_sensitivity}
  \renewcommand{\arraystretch}{1.07}
  \setlength{\tabcolsep}{2.2pt}
  \footnotesize
  \begin{tabular}{@{}llcccccccccc@{}}
    \toprule
    & & \multicolumn{2}{c}{XSTest-h} & \multicolumn{2}{c}{HarmBench} & \multicolumn{2}{c}{Do-Not-Answer} & \multicolumn{2}{c}{OR-Bench} & \multicolumn{2}{c}{XSTest-b} \\
    \cmidrule(lr){3-4}\cmidrule(lr){5-6}\cmidrule(lr){7-8}\cmidrule(lr){9-10}\cmidrule(l){11-12}
    Model                         & $\tau_{\mathrm{refuse}}$ & CRR$\uparrow$ & ASR$\downarrow$ & CRR$\uparrow$ & ASR$\downarrow$ & CRR$\uparrow$ & ASR$\downarrow$ & ORR$\downarrow$ & SHR$\uparrow$ & ORR$\downarrow$ & SHR$\uparrow$ \\
    \midrule
    \multirow{4}{*}{Qwen2.5-3B}   & 0.70                     & \textbf{0.93} & \textbf{0.00}   & \textbf{0.90} & \textbf{0.07}   & \textbf{0.83} & \textbf{0.00}   & 0.26            & 0.51          & 0.15            & 0.83 \\
    & 0.80                     & \textbf{0.93} & \textbf{0.00}   & 0.89          & 0.08            & 0.81          & 0.00            & 0.26            & 0.51          & 0.15            & 0.83 \\
    & 0.90                     & 0.91          & \textbf{0.00}   & 0.87          & 0.08            & 0.79          & 0.00            & \textbf{0.25}   & \textbf{0.51} & \textbf{0.14}   & \textbf{0.84} \\
    & 0.95                     & 0.86          & \textbf{0.00}   & 0.84          & 0.09            & 0.73          & 0.00            & \textbf{0.25}   & \textbf{0.51} & \textbf{0.14}   & \textbf{0.84} \\
    \midrule
    \multirow{4}{*}{Llama-3.2-3B} & 0.70                     & \textbf{0.96} & \textbf{0.01}   & \textbf{0.99} & \textbf{0.01}   & \textbf{0.93} & \textbf{0.00}   & 0.08            & 0.56          & 0.14            & 0.86 \\
    & 0.80                     & 0.92          & 0.02            & 0.98          & 0.02            & 0.90          & 0.01            & 0.07            & 0.57          & \textbf{0.11}   & \textbf{0.89} \\
    & 0.90                     & 0.88          & 0.03            & 0.97          & 0.02            & 0.86          & 0.01            & \textbf{0.05}   & \textbf{0.58} & \textbf{0.11}   & \textbf{0.89} \\
    & 0.95                     & 0.86          & 0.03            & 0.96          & 0.04            & 0.76          & 0.01            & \textbf{0.05}   & \textbf{0.58} & \textbf{0.11}   & \textbf{0.89} \\
    \midrule
    \multirow{4}{*}{Qwen3-4B}     & 0.70                     & \textbf{0.93} & \textbf{0.00}   & \textbf{0.96} & \textbf{0.01}   & \textbf{0.93} & \textbf{0.00}   & 0.54            & 0.40          & 0.15            & 0.85 \\
    & 0.80                     & 0.92          & \textbf{0.00}   & \textbf{0.96} & \textbf{0.01}   & 0.90          & \textbf{0.00}   & 0.53            & 0.41          & 0.14            & 0.86 \\
    & 0.90                     & 0.91          & \textbf{0.00}   & 0.96          & \textbf{0.01}   & 0.88          & \textbf{0.00}   & \textbf{0.53}   & \textbf{0.41} & \textbf{0.13}   & \textbf{0.87} \\
    & 0.95                     & 0.89          & \textbf{0.00}   & 0.96          & \textbf{0.01}   & 0.81          & \textbf{0.00}   & \textbf{0.53}   & \textbf{0.41} & \textbf{0.13}   & \textbf{0.87} \\
    \bottomrule
  \end{tabular}
\end{table*}

As $\tau_{\mathrm{refuse}}$ increases, refusal enforcement is triggered less often. Harmful CRR therefore never increases and generally decreases. Do-Not-Answer is the most sensitive dataset, with CRR reductions of 0.10--0.17 across the three models. The largest reduction occurs for Llama-3.2-3B, from 0.93 to 0.76. HarmBench changes less, and its CRR for Qwen3-4B remains at 0.96.

Gains on benign-sensitive prompts are smaller. Across OR-Bench, the largest reduction is 0.03 for Llama-3.2-3B, and the largest XSTest-b reduction is also 0.03. These gains come with weaker protection on harmful prompts. For example, Qwen2.5-3B HarmBench ASR rises from 0.07 to 0.09. The sweep therefore shows the expected trade-off: a higher refusal threshold can slightly reduce over-refusal, but it can also reduce safety on harmful prompts. This result also explains why a single manually chosen threshold cannot replace the calibrated threshold for each model.

\begin{table*}[!t]
  \centering
  \caption{Generation-stage runtime cost by model. Latency is mean wall-clock time per prompt. $\Delta$ latency reports added milliseconds, with the relative increase in parentheses. Regen. is the regeneration rate, tokens are mean total output tokens, and GPU $\Delta$ is the change in peak allocated memory.}
  \label{tab:runtime_cost}
  \renewcommand{\arraystretch}{1.10}
  \setlength{\tabcolsep}{3.8pt}
  \footnotesize
  \begin{tabular}{@{}lcccccccc@{}}
    \toprule
    Model        & $n$ & Base (ms) & RISA (ms) & $\Delta$ latency & Regen. & Base tokens & RISA tokens & GPU $\Delta$ (MiB) \\
    \midrule
    Qwen2.5-3B   & 400 & 763.76    & 797.26    & +33.50 (+0.04)   & 0.05   & 217.95      & 230.28      & 2.00 \\
    Llama-3.2-3B & 400 & 661.45    & 663.90    & +2.44 (+0.00)    & 0.01   & 212.71      & 215.18      & 0.00 \\
    Qwen3-4B     & 400 & 896.02    & 983.32    & +87.30 (+0.10)   & 0.11   & 196.31      & 223.77      & 0.00 \\
    \bottomrule
  \end{tabular}
\end{table*}

\subsection{Runtime Generation Cost}
\label{subsec:runtime_cost}

Table~\ref{tab:runtime_cost} reports the generation runtime cost for 400 safety prompts per model. The base condition uses a single batched generation pass. For RISA, we reuse the loaded model and the precomputed prompt scores. We make a second generation call only for safety-guided regeneration. This measurement therefore isolates the generation overhead after the prompt scores are available.

Latency overhead scales with the regeneration rate. Qwen3-4B has the highest regeneration rate at 0.11 and the highest relative latency increase at 0.10. Llama-3.2-3B regenerates only 0.01 of prompts, so its added latency is close to zero. Output length follows the same pattern and increases most for Qwen3-4B. Peak GPU memory changes by at most 2 MiB. Safety-guided regeneration is therefore the main source of additional generation cost.

\section{Conclusion}

This paper presents RISA as a runtime verification framework for LLM refusal calibration. Rather than treating refusal behavior as a fixed property of the base model or changing all generations uniformly, RISA inspects the model's initial response, compares its refusal status with a refusal score for the prompt, and decides whether to preserve the response, enforce refusal, or regenerate. The framework combines selective rules with a lightweight hidden-state probe. The rules handle clear prompt patterns, while the probe addresses cases the rules do not resolve. Its main design choice is an asymmetric correction policy. Harmful non-refusals can be replaced with a fixed refusal. Benign over-refusals may be corrected through safety-guided regeneration, but the base model still controls the new answer. Experiments across three instruction-tuned models and multiple safety benchmarks support this design. RISA consistently improves refusal reliability on harmful prompts, while its behavior on benign-sensitive prompts varies across data sources. General utility is largely preserved: GSM8K responses remain unchanged, and MMLU-STEM accuracy losses are small. These results suggest that refusal calibration is better treated as targeted runtime verification with asymmetric action selection than as a single global change in model cautiousness.

\bibliographystyle{IEEEtran}
\bibliography{references}

@inproceedings{ouyang2022training,
  author    = {Long Ouyang and Jeff Wu and Xu Jiang and Diogo Almeida and Carroll L. Wainwright and Pamela Mishkin and others},
  title     = {Training language models to follow instructions with human feedback},
  booktitle = {Advances in Neural Information Processing Systems (NeurIPS)},
  year      = {2022},
  volume    = {35},
  pages     = {27730--27744},
  publisher = {Curran Associates, Inc.}
}

@inproceedings{rafailov2023direct,
  author    = {Rafael Rafailov and Archit Sharma and Eric Mitchell and Christopher D. Manning and Stefano Ermon and Chelsea Finn},
  title     = {Direct Preference Optimization: Your Language Model is Secretly a Reward Model},
  booktitle = {Advances in Neural Information Processing Systems (NeurIPS)},
  year      = {2023},
  volume    = {36},
  pages     = {53728--53741},
  publisher = {Curran Associates, Inc.}
}

@article{ziegler2019fine,
  author    = {Daniel M. Ziegler and Nisan Stiennon and Jeffrey Wu and Tom B. Brown and Alec Radford and Dario Amodei and Paul Christiano and Geoffrey Irving},
  title     = {Fine-Tuning Language Models from Human Preferences},
  journal   = {arXiv preprint arXiv:1909.08593},
  year      = {2019},
  doi       = {10.48550/arXiv.1909.08593},
}

@article{bai2022training,
  author    = {Yuntao Bai and Andy Jones and Kamal Ndousse and Amanda Askell and Anna Chen and Nova DasSarma and others},
  title     = {Training a Helpful and Harmless Assistant with Reinforcement Learning from Human Feedback},
  journal   = {arXiv preprint arXiv:2204.05862},
  year      = {2022},
  doi       = {10.48550/arXiv.2204.05862},
}

@article{askell2021general,
  author    = {Amanda Askell and Yuntao Bai and Anna Chen and Dawn Drain and Deep Ganguli and Tom Henighan and others},
  title     = {A General Language Assistant as a Laboratory for Alignment},
  journal   = {arXiv preprint arXiv:2112.00861},
  year      = {2021},
  doi       = {10.48550/arXiv.2112.00861},
}

@inproceedings{bianchi2024safetunedllamas,
  author    = {Federico Bianchi and Mirac Suzgun and Giuseppe Attanasio and Paul R{\"o}ttger and Dan Jurafsky and Tatsunori Hashimoto and James Zou},
  title     = {Safety-Tuned {LLaMAs}: Lessons From Improving the Safety of Large Language Models that Follow Instructions},
  booktitle = {The Twelfth International Conference on Learning Representations},
  year      = {2024},
}

@inproceedings{dai2024saferlhf,
  author    = {Josef Dai and Xuehai Pan and Ruiyang Sun and Jiaming Ji and Xinbo Xu and Mickel Liu and Yizhou Wang and Yaodong Yang},
  title     = {Safe {RLHF}: Safe Reinforcement Learning from Human Feedback},
  booktitle = {The Twelfth International Conference on Learning Representations},
  year      = {2024}
}

@inproceedings{sun2024salmon,
  author    = {Zhiqing Sun and Yikang Shen and Hongxin Zhang and Qinhong Zhou and Zhenfang Chen and David Cox and Yiming Yang and Chuang Gan},
  title     = {{SALMON}: Self-Alignment with Instructable Reward Models},
  booktitle = {The Twelfth International Conference on Learning Representations},
  year      = {2024},
}

@inproceedings{dabas2025just,
  author    = {Mahavir Dabas and Si Chen and Charles Fleming and Ming Jin and Ruoxi Jia},
  title     = {Just Enough Shifts: Mitigating Over-Refusal in Aligned Language Models with Targeted Representation Fine-Tuning},
  booktitle = {Proceedings of the 42nd International Conference on Machine Learning},
  year      = {2025},
  volume    = {267},
  series    = {Proceedings of Machine Learning Research},
  pages     = {11846--11861},
  month     = {13--19 Jul},
  publisher = {PMLR},
}

@article{zou2023representation,
  author    = {Andy Zou and Long Phan and Sarah Chen and James Campbell and Phillip Guo and Richard Ren and others},
  title     = {Representation Engineering: A Top-Down Approach to {AI} Transparency},
  journal   = {arXiv preprint arXiv:2310.01405},
  year      = {2023},
  doi       = {10.48550/arXiv.2310.01405},
}

@article{turner2023activation,
  author    = {Alexander Matt Turner and Lisa Thiergart and Gavin Leech and David Udell and Juan J. Vazquez and Ulisse Mini and Monte MacDiarmid},
  title     = {Steering Language Models With Activation Engineering},
  journal   = {arXiv preprint arXiv:2308.10248},
  year      = {2023},
  doi       = {10.48550/arXiv.2308.10248},
}

@inproceedings{zhao2025harmfulnessrefusal,
  author    = {Jiachen Zhao and Jing Huang and Zhengxuan Wu and David Bau and Weiyan Shi},
  title     = {{LLMs} Encode Harmfulness and Refusal Separately},
  booktitle = {The Thirty-Ninth Annual Conference on Neural Information Processing Systems},
  year      = {2025},
}

@article{inan2023llama,
  author    = {Hakan Inan and Kartikeya Upasani and Jianfeng Chi and Rashi Rungta and Krithika Iyer and Yuning Mao and Michael Tontchev and Qing Hu and Brian Fuller and Davide Testuggine and Madian Khabsa},
  title     = {{Llama Guard}: {LLM}-based Input-Output Safeguard for Human-{AI} Conversations},
  journal   = {arXiv preprint arXiv:2312.06674},
  year      = {2023},
  doi       = {10.48550/arXiv.2312.06674},
}

@misc{openai2026moderation,
  author       = {{OpenAI}},
  title        = {Moderation},
  howpublished = {OpenAI API documentation},
  year         = {2026},
  note         = {Accessed: 2026-07-26},
}

@inproceedings{cui2025orbench,
  author    = {Justin Cui and Wei-Lin Chiang and Ion Stoica and Cho-Jui Hsieh},
  title     = {{OR-Bench}: An Over-Refusal Benchmark for Large Language Models},
  booktitle = {Proceedings of the 42nd International Conference on Machine Learning},
  year      = {2025},
  volume    = {267},
  series    = {Proceedings of Machine Learning Research},
  pages     = {11515--11542},
  month     = {13--19 Jul},
  publisher = {PMLR},
}

@inproceedings{rottger2024xstest,
  author    = {Paul R{"o}ttger and Hannah Kirk and Bertie Vidgen and Giuseppe Attanasio and Federico Bianchi and Dirk Hovy},
  title     = {{XST}est: A Test Suite for Identifying Exaggerated Safety Behaviours in Large Language Models},
  booktitle = {Proceedings of the 2024 Conference of the North American Chapter of the Association for Computational Linguistics: Human Language Technologies (Volume 1: Long Papers)},
  month     = jun,
  year      = {2024},
  address   = {Mexico City, Mexico},
  pages     = {5377--5400},
  publisher = {Association for Computational Linguistics},
  doi       = {10.18653/v1/2024.naacl-long.301}
}

@inproceedings{mazeika2024harmbench,
  author    = {Mantas Mazeika and Long Phan and Xuwang Yin and Andy Zou and Zifan Wang and Norman Mu and Elham Sakhaee and Nathaniel Li and Steven Basart and Bo Li and David Forsyth and Dan Hendrycks},
  title     = {{HarmBench}: A Standardized Evaluation Framework for Automated Red Teaming and Robust Refusal},
  booktitle = {Proceedings of the 41st International Conference on Machine Learning},
  year      = {2024},
  volume    = {235},
  series    = {Proceedings of Machine Learning Research},
  pages     = {35181--35224},
  month     = {21--27 Jul},
  publisher = {PMLR}
}

@inproceedings{wang2024donotanswer,
  author    = {Yuxia Wang and Haonan Li and Xudong Han and Preslav Nakov and Timothy Baldwin},
  title     = {Do-Not-Answer: Evaluating Safeguards in {LLM}s},
  booktitle = {Findings of the Association for Computational Linguistics: EACL 2024},
  month     = mar,
  year      = {2024},
  address   = {St. Julian's, Malta},
  pages     = {896--911},
  publisher = {Association for Computational Linguistics},
  doi       = {10.18653/v1/2024.findings-eacl.61}
}

@inproceedings{ji2023beavertails,
  author    = {Jiaming Ji and Mickel Liu and Josef Dai and Xuehai Pan and Chi Zhang and Ce Bian and Boyuan Chen and Ruiyang Sun and Yizhou Wang and Yaodong Yang},
  title     = {{BeaverTails}: Towards Improved Safety Alignment of {LLM} via a Human-Preference Dataset},
  booktitle = {Advances in Neural Information Processing Systems (NeurIPS)},
  year      = {2023},
  volume    = {36},
  pages     = {24678--24704},
  publisher = {Curran Associates, Inc.}
}

@article{touvron2023llama2,
  author    = {Hugo Touvron and Louis Martin and Kevin Stone and Peter Albert and Amjad Almahairi and Yasmine Babaei and others},
  title     = {{Llama 2}: Open Foundation and Fine-Tuned Chat Models},
  journal   = {arXiv preprint arXiv:2307.09288},
  year      = {2023},
  doi       = {10.48550/arXiv.2307.09288},
}

@inproceedings{meade2023using,
  author    = {Nicholas Meade and Spandana Gella and Devamanyu Hazarika and Prakhar Gupta and Di Jin and Siva Reddy and Yang Liu and Dilek Hakkani-Tur},
  title     = {Using In-Context Learning to Improve Dialogue Safety},
  booktitle = {Findings of the Association for Computational Linguistics: EMNLP 2023},
  month     = dec,
  year      = {2023},
  address   = {Singapore},
  pages     = {11882--11910},
  publisher = {Association for Computational Linguistics},
  doi       = {10.18653/v1/2023.findings-emnlp.796}
}

@inproceedings{xu2024safedecoding,
  author    = {Zhangchen Xu and Fengqing Jiang and Luyao Niu and Jinyuan Jia and Bill Yuchen Lin and Radha Poovendran},
  title     = {{SafeDecoding}: Defending against Jailbreak Attacks via Safety-Aware Decoding},
  booktitle = {Proceedings of the 62nd Annual Meeting of the Association for Computational Linguistics (Volume 1: Long Papers)},
  month     = aug,
  year      = {2024},
  address   = {Bangkok, Thailand},
  pages     = {5587--5605},
  publisher = {Association for Computational Linguistics},
  doi       = {10.18653/v1/2024.acl-long.303}
}

@inproceedings{banerjee2025safeinfer,
  author    = {Somnath Banerjee and Sayan Layek and Soham Tripathy and Shanu Kumar and Animesh Mukherjee and Rima Hazra},
  title     = {{SafeInfer}: Context Adaptive Decoding Time Safety Alignment for Large Language Models},
  booktitle = {Proceedings of the AAAI Conference on Artificial Intelligence},
  year      = {2025},
  volume    = {39},
  number    = {26},
  pages     = {27188--27196},
  doi       = {10.1609/aaai.v39i26.34927}
}

@inproceedings{brown2020language,
  author    = {Tom B. Brown and Benjamin Mann and Nick Ryder and Melanie Subbiah and Jared Kaplan and Prafulla Dhariwal and others},
  title     = {Language Models are Few-Shot Learners},
  booktitle = {Advances in Neural Information Processing Systems (NeurIPS)},
  year      = {2020},
  volume    = {33},
  pages     = {1877--1901},
  publisher = {Curran Associates, Inc.}
}

@inproceedings{christiano2017deep,
  author    = {Paul F. Christiano and Jan Leike and Tom B. Brown and Miljan Martic and Shane Legg and Dario Amodei},
  title     = {Deep Reinforcement Learning from Human Preferences},
  booktitle = {Advances in Neural Information Processing Systems (NeurIPS)},
  year      = {2017},
  volume    = {30},
  pages     = {4299--4307},
  publisher = {Curran Associates, Inc.}
}

@article{bai2022constitutional,
  author    = {Yuntao Bai and Saurav Kadavath and Sandipan Kundu and Amanda Askell and Jackson Kernion and Andy Jones and others},
  title     = {Constitutional {AI}: Harmlessness from {AI} Feedback},
  journal   = {arXiv preprint arXiv:2212.08073},
  year      = {2022},
  doi       = {10.48550/arXiv.2212.08073},
}

@inproceedings{rimsky2024steering,
  author    = {Nina Rimsky and Nick Gabrieli and Julian Schulz and Meg Tong and Evan Hubinger and Alexander Turner},
  title     = {Steering {Llama 2} via Contrastive Activation Addition},
  booktitle = {Proceedings of the 62nd Annual Meeting of the Association for Computational Linguistics (Volume 1: Long Papers)},
  month     = aug,
  year      = {2024},
  address   = {Bangkok, Thailand},
  pages     = {15504--15522},
  publisher = {Association for Computational Linguistics},
  doi       = {10.18653/v1/2024.acl-long.828},
}

@inproceedings{arditi2024refusal,
  author    = {Andy Arditi and Oscar Obeso and Aaquib Syed and Daniel Paleka and Nina Panickssery and Wes Gurnee and Neel Nanda},
  title     = {Refusal in Language Models Is Mediated by a Single Direction},
  booktitle = {Advances in Neural Information Processing Systems},
  year      = {2024},
  volume    = {37},
  pages     = {136037--136083},
  publisher = {Curran Associates, Inc.}
}

@inproceedings{qi2026adacd,
  author    = {Yupeng Qi and Ziyu Lyu and Lixin Cui and Lu Bai and Feng Xia},
  title     = {Please Refuse to Answer Me! Mitigating Over-Refusal in Large Language Models via Adaptive Contrastive Decoding},
  booktitle = {Proceedings of the 64th Annual Meeting of the Association for Computational Linguistics (Volume 1: Long Papers)},
  month     = jul,
  year      = {2026},
  address   = {San Diego, California, United States},
  pages     = {39308--39325},
  publisher = {Association for Computational Linguistics},
  doi       = {10.18653/v1/2026.acl-long.1823}
}

@inproceedings{shi2024navigating,
  author    = {Chenyu Shi and Xiao Wang and Qiming Ge and Songyang Gao and Xianjun Yang and Tao Gui and Qi Zhang and Xuanjing Huang and Xun Zhao and Dahua Lin},
  title     = {Navigating the OverKill in Large Language Models},
  booktitle = {Proceedings of the 62nd Annual Meeting of the Association for Computational Linguistics (Volume 1: Long Papers)},
  month     = aug,
  year      = {2024},
  address   = {Bangkok, Thailand},
  pages     = {4602--4614},
  publisher = {Association for Computational Linguistics},
  doi       = {10.18653/v1/2024.acl-long.253}
}

@inproceedings{rebedea2023nemo,
  author    = {Traian Rebedea and Razvan Dinu and Makesh Narsimhan Sreedhar and Christopher Parisien and Jonathan Cohen},
  title     = {{NeMo} Guardrails: {A} Toolkit for Controllable and Safe {LLM} Applications with Programmable Rails},
  booktitle = {Proceedings of the 2023 Conference on Empirical Methods in Natural Language Processing: System Demonstrations},
  month     = dec,
  year      = {2023},
  address   = {Singapore},
  pages     = {431--445},
  publisher = {Association for Computational Linguistics},
  doi       = {10.18653/v1/2023.emnlp-demo.40}
}

\end{document}